\documentclass[12pt]{article}
\usepackage{amsfonts}
\usepackage{amssymb}
\usepackage{amsbsy}
\usepackage{graphics}

\input epsf.sty

\newlength{\TZ}
\newcommand{\BEQ}{\begin{equation}}          
\newcommand{\BEA}{\begin{eqnarray}}
\newcommand{\EEQ}{\end{equation}}            
\newcommand{\EEA}{\end{eqnarray}}
\newcommand{\eps}{\varepsilon}               
\newcommand{\D}{{\rm d}}                     
\newcommand{\II}{{\rm i}}                    
\newcommand{\wit}[1]{\widetilde{#1}}         

\renewcommand{\vec}[1]{\boldsymbol{#1}}      

\newcommand{\zeile}[1]{\vskip #1 \baselineskip} 

\catcode`\@=11
\def\numberbysection{\@addtoreset{equation}{section}
                     \def\theequation{\thesection.\arabic{equation}}}

\def\up#1{\raise 1ex\hbox{\sevenrm#1}}
\def\build#1_#2^#3{\mathrel{\mathop{\kern 0pt#1}\limits_{#2}^{#3}}}

\numberbysection

\begin{document}

\begin{titlepage}

\vskip 1.5 cm
\begin{center}
{\Large \bf Dynamical symmetries of semi-linear Schr\"odinger and diffusion equations}
\end{center}

\vskip 2.0 cm
\centerline{ {\bf Stoimen Stoimenov}$^{a,b}$ and {\bf Malte Henkel}$^{a}$ }
\vskip 0.5 cm
\centerline {$^a$Laboratoire de Physique des
Mat\'eriaux,\footnote{Laboratoire associ\'e au CNRS UMR 7556}
Universit\'e Henri Poincar\'e Nancy I,}
\centerline{ B.P. 239,
F -- 54506 Vand{\oe}uvre l\`es Nancy Cedex, France}
\centerline{$^b$Institute of Nuclear Research and Nuclear Energy,
Bulgarian Academy of Sciences,}
\centerline{1784 Sofia, Bulgaria}
\begin{abstract}
Conditional and Lie symmetries of semi-linear $1D$ Schr\"odinger and diffusion 
equations are studied if the mass (or the diffusion constant) is considered 
as an additional variable. In this way, dynamical symmetries of
semi-linear Schr\"odinger equations become related to the parabolic
and almost-parabolic subalgebras of a three-dimensional conformal Lie algebra
$(\mathfrak{conf}_3)_{\mathbb{C}}$. 
We consider non-hermitian representations and also include a dimensionful
coupling constant of the non-linearity. The corresponding representations
of the parabolic and almost-parabolic subalgebras of
$(\mathfrak{conf}_3)_{\mathbb{C}}$ are classified and the
complete list of conditionally invariant 
semi-linear Schr\"odinger equations is obtained. 
Possible applications to the dynamical scaling behaviour of phase-ordering
kinetics are discussed.  
\end{abstract}
\zeile{10}


\end{titlepage}

\section{Introduction}

Symmetries have since a long time played a major r\^ole in physics. 
For example, starting with Lie in 1881, the maximal kinematic invariance
group of the free diffusion equation, where $\cal M$ is a constant 
\BEQ \label{eq:diffGl}
2{\cal M}\partial_t\Phi-\partial^2_{\vec r}\Phi =0
\EEQ
has been studied, which is the so-called Schr\"odinger group {\sl Sch}$(d)$
\cite{Nied72} which also arises a dynamical symmetry of the non-relativistic
free particle \cite{Kast68,Hage72,Baru80}. 
We shall denote its Lie algebra by $\mathfrak{sch}_d$. For $d=1$ space
dimensions, $\mathfrak{sch}_1$ is spanned by the generators
\BEA
X_{-1} &=& -\partial_t \;\; , \;\; Y_{-{1\over 2}} \:=\: -\partial_r
\nonumber \\
Y_{1\over 2} &=&-t\partial_r-{\cal M} r
\nonumber \\
X_0 &=&-t\partial_t-{1\over 2}r\partial_r-{x\over 2}
\label{eq:sch1} \\
X_1 &=& -t^2\partial_t-tr\partial_r-{{\cal M}\over 2}r^2-xt
\nonumber \\
M_0 &=&-{\cal M}
\nonumber
\EEA
which allows to write the non-vanishing commutators compactly as
$[X_n,X_{n'}]=(n-n')X_{n+n'}$, $[X_n,Y_m]=(n/2-m)Y_{n+m}$,
$[Y_{\frac{1}{2}},Y_{-\frac{1}{2}}]=M_0$ where $n,n'\in\{\pm 1,0\}$ and
$m\in\{\pm\frac{1}{2}\}$ (see \cite{Henk02} for generalizations to $d>1$). 
The free Schr\"odinger equation is recovered from the analytical continuation
${\cal M}=\II m$. It is a well-known fact that the same
group also acts as kinematic invariance group of certain non-linear
Schr\"odinger equations of the form  
\BEQ \label{eq:NLSE}
2m\II\partial_t\Phi-\partial^2_{\vec r}\Phi =F(t,r,\Phi , \Phi^*)
\EEQ
If the potential is chosen in the form $F=c\left(\Phi\Phi^*\right)^{2/d}\Phi$, 
where $c$ is a constant, then eq.~(\ref{eq:NLSE}) is invariant under 
$\mathfrak{sch}_d$ provided the scaling dimension of $\Phi$ is taken to be
$x=d/2$, see e.g. \cite{Boye76,Fush93,Ride93}. Nonlinear Schr\"odinger equations
arise in many physical applications, for recent reviews see 
\cite{Aran02,Sule99}. We also mention the recently established 
Schr\"odinger-invariance of non-relativistic fluid dynamics 
\cite{Hass00,ORai01}. 
Mathematically, there has been a lot of effort to establish the existence of
solutions with certain regularity properties, see e.g. \cite{Bour00}, 
and on the other hand the group classification of non-linear Schr\"odinger
equations has been intensively studied, see \cite{Boye76,Ride93,Fush93,Fush95,Guen99,Cher00,Cher03,Popo04,Cher04} 
and references therein. Sprectral properties are studied in \cite{Bazh04}. A great deal is known about the representations of 
$\mathfrak{sch}_d$ \cite{Perr77,Medi85,Doeb95,Dobr97,Fein04} and this can be applied to find symbolic solutions of non-linear Schr\"odinger 
equations.\footnote{Related questions
include the dynamical symmetries in de Sitter space \cite{Dira63,Fron65},
$q$-deformations \cite{Flor94}, Chern-Simons theory \cite{Dunn89,deMo99}, 
difference equations \cite{Dobr99,Levi05} or even the integrable
quantum non-linear Schr\"odinge equation, see e.g. \cite{Caud04}.} 

Very similar equations have been studied in attempts to understand the
coarsening process which systems undergo after having been quenched from a
disordered initial state to below their critical point, see e.g. 
\cite{Bray94,Bouc00,Bray00,Cugl02,Godr02,Cris03,Henk04,Henk05}
for reviews. In its most simple setting, one considers a ferromagnetic system 
(e.g. described by an Ising model) which from some initial high-temperature 
state is rapidly 
quenched into its ordered phase below its critical temperature $T_c>0$ where 
there are at least two competing equilbrium states. 
Although the system relaxes locally towards one of the equilibrium states, a
{\em global} relaxation is not possible and this leads to a very slow evolution
of the macroscopic observables. From a miscrocopic point of view, the system's
evolution is characterized by the formation of correlated domains of 
time-dependent linear size $L(t)$ and one typically finds a power-law
behaviour $L(t)~\sim t^{1/z}$ where $z$ is called the {\em dynamical exponent}. 
This in turn signals a dynamical scale-invariance in the system's evolution. 
A coarse-grained description is usually given in terms of the order parameter
$\Phi=\Phi(t,\vec{r})$, which in the absence of any macroscopic conservation
law is assumed to satisfy \cite{Hohe77,Bray94}
\BEQ \label{modelA}
\frac{\partial \Phi}{\partial t} = \Gamma \vec{\nabla}^2 \Phi 
-\frac{\D V(\Phi)}{\D \Phi} 
\EEQ
where $\Gamma$ is a kinetic coefficient and $V(\Phi)$ is the potential which 
enters into the Ginzburg-Landau free-energy functional 
and which is assumed to have a double-well structure, i.e. 
$V(\Phi)=(1-\Phi^2)^2$ \cite{Bray94,Bray00}. For simplicity, we dropped
here the thermal noise which is known \cite{Bray94} to be irrelevant for the 
long-time behaviour we are interested in. 
The disordered initial state enters through `white-noise' initial conditions
and one looks for the long-time behaviour of the solutions of (\ref{modelA}). 

It has turned out to be convenient to characterize this
evolution in terms of the two-time autocorrelation $C(t,s)$ and the 
associated two-time autoresponse $R(t,s)$ defined as 
\BEQ
C(t,s) :=\langle \Phi(t) \Phi (s)\rangle \;\; , \;\;
R(t,s) := \left.{\delta\langle \Phi (t)\rangle \over \delta h(s)}\right|_{h=0}
\EEQ
where $h(s)$ is the time-dependent magnetic field conjugate to $\Phi$ and
$\langle .\rangle$ denotes the average over the
fluctuations in the initial state (and thermal histories). Here, $t$
is referred to as {\em observation time} and $s$ as {\em waiting time}.
By definition, the system undergoes {\em ageing} if $C$ or $R$
depend on both $t$ and $s$ and not merely on the difference $\tau=t-s$. 
It is well-accepted (although still unproven) that in the ageing regime, 
that is for times $t,s\gg t_{\rm micro}$ and $t-s\gg t_{\rm micro}$,
where $t_{\rm micro}$ is some microscopic time scale, dynamical scaling 
holds true such that 
\BEQ
C(t,s)=s^{-b}f_C({t/s}) \;\; , \;\;
R(t,s)=s^{-1-a}f_R({t/s})
\EEQ
and one would like to be able to compute the scaling functions
$f_{C,R}(y)$. The free diffusion equation is the simplest example of a
system undergoing ageing \cite{Cugl94}. Motivated by a formal analogy with
the well-known conformal invariance in equilibrium critical phenomena, it
has been suggested that the Schr\"odinger group might play a similar r\^ole
in certain ageing systems \cite{Henk94,Henk02}.\footnote{Here we 
only consider the phase-ordering kinetics of ferromagnetic systems 
without any macroscopic conservation laws and quenched to below their 
critical point. Then $z=2$ has been derived \cite{Rute95} and a necessary
condition for the applicability of Schr\"odinger-invariance is satisfied.} 
If this working hypothesis is made, explicit forms for the scaling
function $f_R(y)$ \cite{Henk01,Henk02} and more recently also for $f_C(y)$
\cite{Pico04,Henk04b} can been derived. These agree with the exact results
in several soluble models, such as the spherical model
\cite{Godr00b,Zipp00,Corb02,Pico02}, the $1D$ Glauber-Ising model 
\cite{Godr00a,Lipp00,Henk03d}, 
the critical voter model \cite{Dorn01,Dorn02}, 
the bosonic versions of the contact and the
pair-contact processes \cite{Baum05} and the free random walk \cite{Cugl94}. 
Furthermore, these forms 
also agree with the results of numerical simulations in the $2D$ and $3D$ kinetic Ising \cite{Henk03c} and XY models \cite{Abri04a,Abri04b} and with 
recent results in the $2D$ three-states Potts model \cite{Lore05}.

In order to understand whether these observations may be viewed as manifestations of a dynamical 
symmetry we remark the following: 
\begin{enumerate}
\item In ageing phenomena, time-translation invariance is broken. Therefore,
one should {\it a priori} not expect full Schr\"odinger-invariance 
but at best invariance under
some sub-algebra of $\mathfrak{sch}_d$ which does not contain 
time-translations. For a classification of the Lie subalgebras of 
$\mathfrak{sch}_1$ see \cite{Boye76}.
\item If one considers equations as (\ref{modelA}) as the classical equations
of motion of a field-theory, sufficient conditions for the validity
of Schr\"odinger-invariance can be formulated. If that field-theory is
local and taking the analogous case of
conformal invariance as a guide, one may show from a consideration of the
energy-momentum tensor that the special Schr\"odinger-invariance 
(generated by $X_1$) follows provided that dynamical scaling and 
in addition Galilei-invariance are satisfied \cite{Henk03b}. 
\item If the deterministic part of the field-theory is Galilei-invariant, then both $C(t,s)$ and $R(t,s)$ in the presence of noise can be expressed
in terms of quantities calculable from the noiseless part \cite{Pico04}.
Hence it is enough to study the symmetries of the deterministic part, which
reduces the problem to understanding the symmetries, especially the Galilei-invariance, of a non-linear partial differential equation, such as
(\ref{modelA}).  
\end{enumerate}
At first sight, the problem of finding the Galilei-invariant semi-linear Schr\"odinger equations (\ref{eq:NLSE}) 
seems to have been answered long ago and 
only allows the specific potential $F$ quoted above \cite{Fush93}.
Furthermore, complex-valued solutions $\phi$ of eq.~(\ref{eq:NLSE})
are required, which apparently excludes applications to kinetic equations such
as (\ref{modelA}) with real-valued solutions. 
In this paper, we shall propose a way around these difficulties.

\begin{figure}[t]
\centerline{\epsfxsize=1.5in\ \epsfbox{
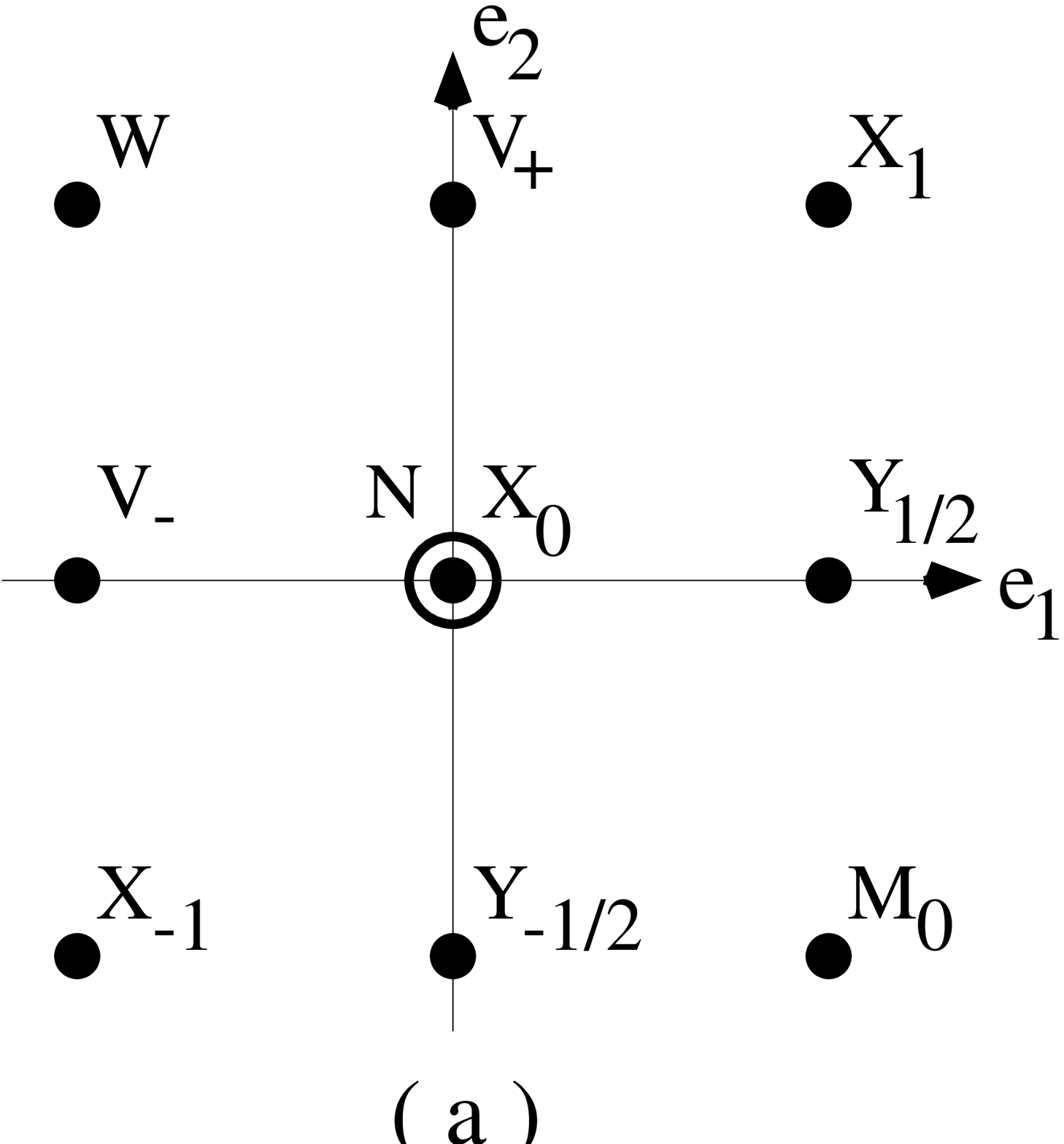} ~
\epsfxsize=1.5in\epsfbox{
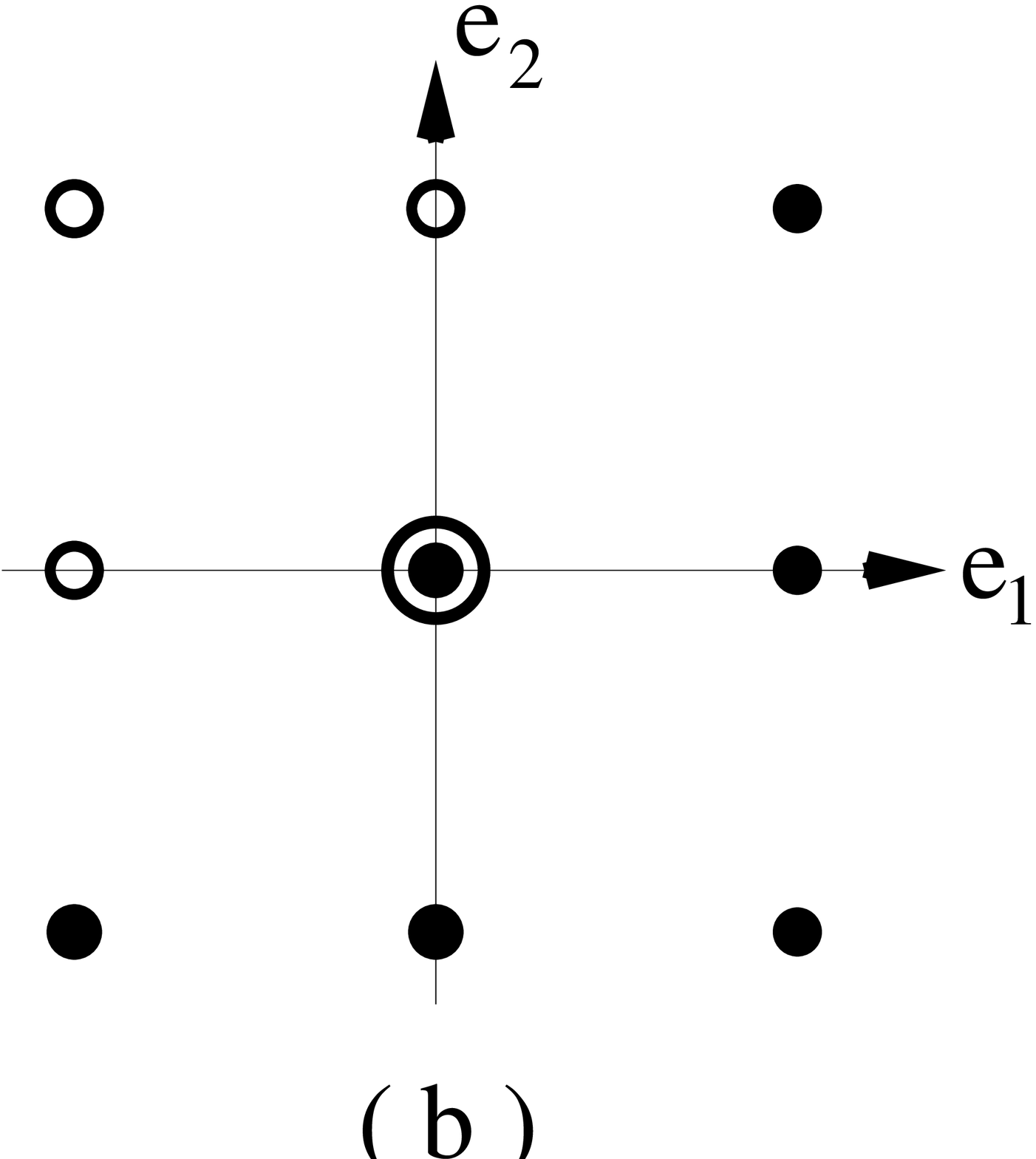} ~
\epsfxsize=1.5in\epsfbox{
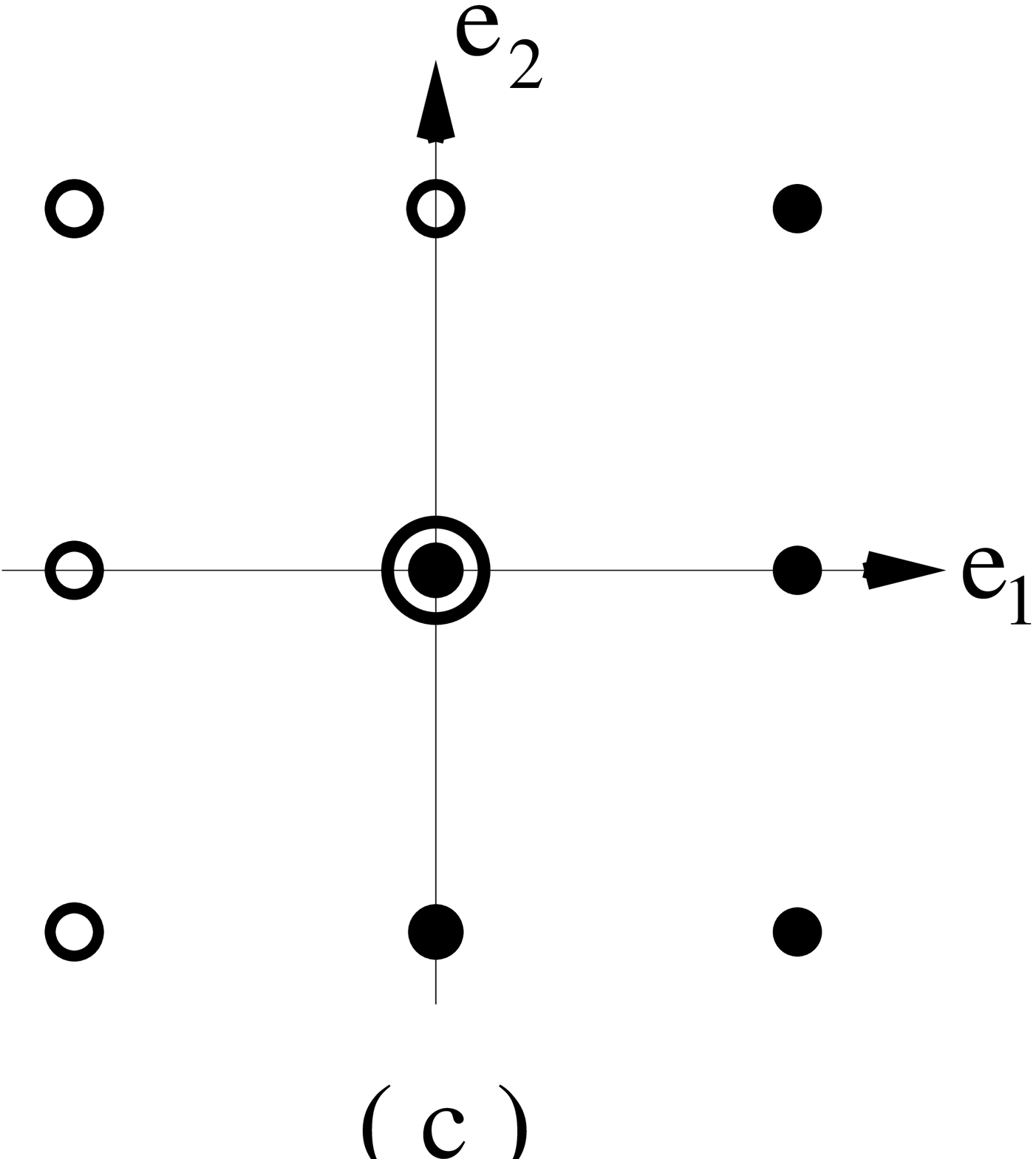} ~
\epsfxsize=1.5in\epsfbox{
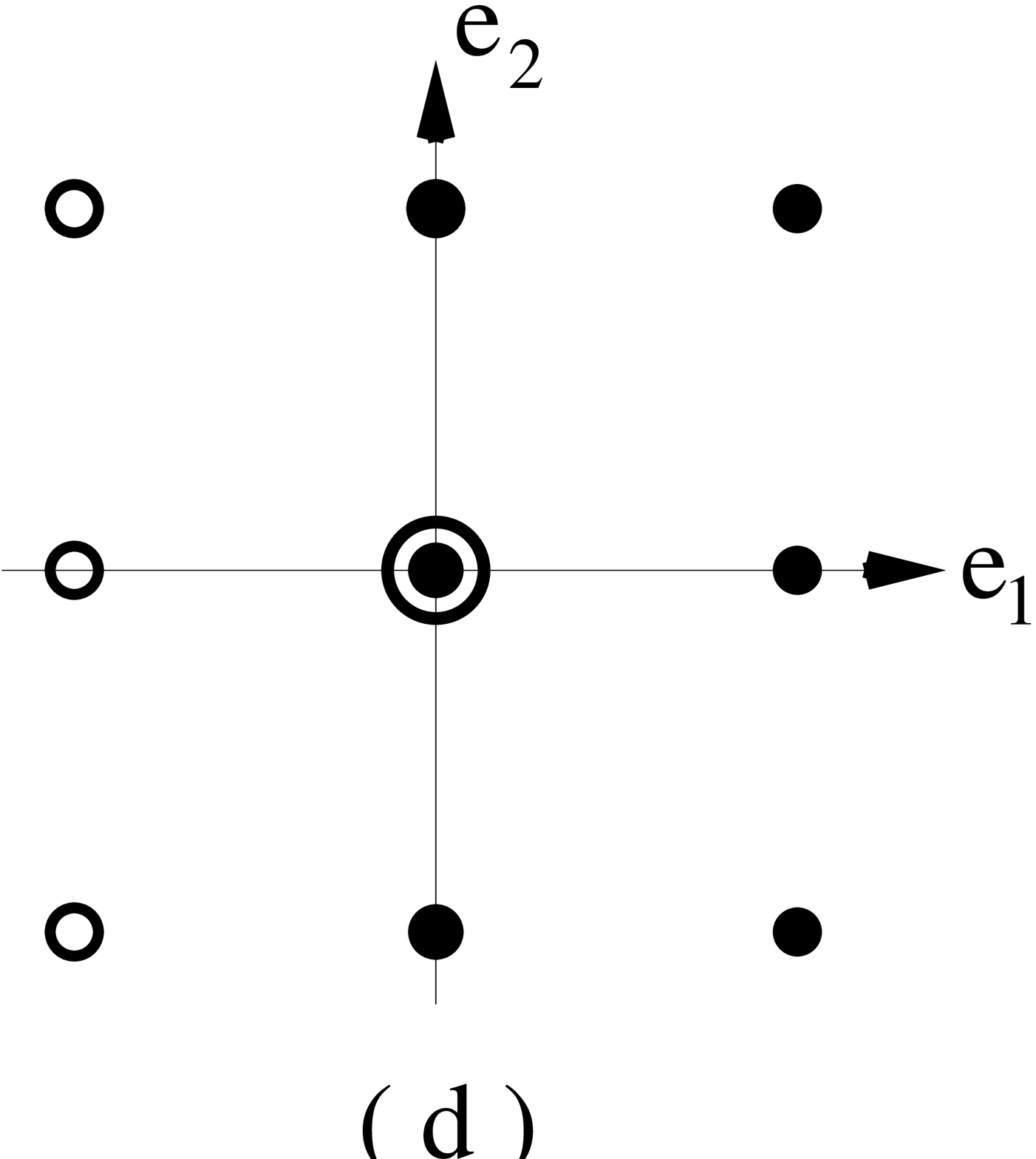}
}
\caption[Root space]{(a) Root diagram of the complex Lie algebra $B_2$ and the 
identification of the generators (\ref{eq:zetaGen},\ref{eq:zetaGen2}) 
of the complexified conformal Lie algebra
$(\mathfrak{conf}_3)_{\mathbb{C}}\supset(\mathfrak{sch}_1)_{\mathbb{C}}$.
The double circle in the center denotes the Cartan subalgebra.   
The generators belonging to the three non-isomorphic parabolic subalgebras 
\protect{\cite{Henk03b}} 
are indicated by the full points, namely
(b) $\wit{\mathfrak{sch}}_1$, (c) $\wit{\mathfrak{age}}_1$ and
(d) $\wit{\mathfrak{alt}}_1$. 
\label{Bild1}}
\end{figure}

We begin by recalling that
the Schr\"odinger algebra $\mathfrak{sch}_d$ in $d$ spatial dimensions 
is a subalgebra of the complexified conformal algebra
$(\mathfrak{conf}_{d+2})_{\mathbb{C}}$ in $d+2$ dimensions 
\cite{Burd73,Henk03b}. A simple way
of seeing this is to treat the `mass' ${\cal M}=\II m$ as a further dynamical
variable and to introduce a new wave function \cite{Giul96,Henk03b}
\BEQ \label{eq:Fourier}
\Phi= \Phi_m(t,\vec{r})=
{1\over \sqrt{2\pi\,}}\int_{\mathbb{R}}\!\D\zeta\:
e^{-\II m\zeta}\Psi(\zeta,t,\vec{r})
\EEQ
For notational simplicity, we restrict from now on to the case $d=1$. Then
the generators (\ref{eq:sch1}) become 
\BEA
X_{-1} &=& -\partial_t \;\; , \;\; Y_{-{1\over 2}}=-\partial_r
\nonumber \\
Y_{1\over 2} &=& -t\partial_r-r\partial_{\zeta } \nonumber \\
X_0 &=& -t\partial_t-{1\over 2}r\partial_r-{x\over 2} 
\label{eq:zetaGen} \\
X_1 &=& -t^2\partial_t-tr\partial_r-{1\over 2}r^2\partial_{\zeta}-xt
\nonumber \\
M_0 &=& -\partial_{\zeta} \nonumber
\EEA
The free Schr\"odinger equation (\ref{eq:diffGl}) then becomes
$\left(2\partial_{\zeta}\partial_t -\partial_r^2\right)\Psi=0$ which through
a further change of variables can be brought into 
a massless Klein-Gordon/Laplace 
equation in three dimensions which has the simple Lie algebra 
$(\mathfrak{conf}_3)_{\mathbb{C}}\cong\mathfrak{so}(5,\mathbb{C})\cong B_2$ 
as dynamical symmetry. It is useful to illustrate this in terms of a root 
diagram, see figure~\ref{Bild1}a, from which the correspondence between the
roots and the generators of $\mathfrak{sch}_1$ can be read off. 
Four additional generators should be added in order to
get the full conformal algebra $(\mathfrak{conf}_3)_{\mathbb{C}}$
which we take in the form \cite{Henk03b}

\BEA
N &=& -t\partial_t+\zeta\partial_{\zeta} \nonumber \\
V_- &=& -\zeta \partial_r-r\partial_t \nonumber \\
W &=& -\zeta^2\partial_{\zeta}-\zeta r\partial_r
-{1\over 2}r^2\partial_t-x\zeta
\label{eq:zetaGen2} \\
V_+ &=& -2tr\partial_t-2\zeta r\partial_{\zeta }-
(r^2+2\zeta t)\partial_r-2xr.
\nonumber
\EEA
In applications to ageing phenomena, it turned out that the parabolic
subalgebras of $\mathfrak{conf}_3$ play an important r\^ole.\footnote{The 
minimal standard parabolic subalgebra $\mathfrak{s}_0$ of a simple complex Lie 
algebra $\mathfrak{g}$ is spanned by the Cartan subalgebra $\mathfrak{h}$ of 
$\mathfrak{g}$ and the set of positive roots. A standard parabolic subalgebra
$\mathfrak{s}\subset \mathfrak{g}$ is a subalgebra of $\mathfrak{g}$ which 
contains $\mathfrak{s}_0$ \cite{Knap86}.} The complete list of non-isomorphic
parabolic subalgebras of $\mathfrak{conf}_3$ is as follows \cite{Henk03b}
\begin{enumerate}
\item $\wit{\mathfrak{sch}}_1$, spanned by the set
$\{ X_{-1,0,1}, Y_{-\frac{1}{2},\frac{1}{2}}, M_0, N\}$. 
\item $\wit{\mathfrak{age}}_1$, spanned by the set
$\{ X_{0,1}, Y_{-{1\over 2},{1\over 2}}, M_0, N\}$. 
\item $\wit{\mathfrak{alt}}_1$, spanned by the set 
$\{ D,X_1,Y_{-{1\over 2},{1\over 2}},M_0,N,V_+\}$.
\end{enumerate}
Here we used the generator $D$ of the full dilatations
\BEQ\label{eq:ddef}
D := 2 X_0 - N = -t\partial_t-r\partial_r-\zeta \partial_{\zeta }-x 
\EEQ
In figure~\ref{Bild1}bcd, we illustrate the definition of these three 
parabolic subalgebras through their root diagrams. 

One can also consider the corresponding subalgebras
without the generator $N$, namely
\begin{enumerate}
\item ${\mathfrak{sch}}_1$, spanned by the set
$\{ X_{-1,0,1}, Y_{-\frac{1}{2},\frac{1}{2}}, M_0\}$. 
\item ${\mathfrak{age}}_1$, spanned by the set 
$\{ X_{0,1}, Y_{-{1\over 2},{1\over 2}}, M_0\}$. 
\item ${\mathfrak{alt}}_1$, spanned by the set 
$\{ D,X_1,Y_{-{1\over 2},{1\over 2}},M_0,V_+\}$.
\end{enumerate}
We call {\em almost-parabolic subalgebras} those subalgebras of a
parabolic subalgebra $\mathfrak{s}$ which are obtained by leaving out one
of the generators of the Cartan subalgebra $\mathfrak{h}$ of $\mathfrak{s}$. 
Therefore, $\mathfrak{sch}_1,\mathfrak{age}_1$ and $\mathfrak{alt}_1$ are almost parabolic, because we merely have to add the generator 
$N\in \mathfrak{h}$ in order to make them parabolic subalgebras. 
In view of possible applications to ageing phenomena, we shall be mainly 
interested in the algebras\footnote{The name of $\mathfrak{age}_1$ comes of 
course from {\em ageing}, while $\mathfrak{alt}_1$ is inspired by its German 
equivalent, {\em altern}.} $\mathfrak{age}_1$ and $\mathfrak{alt}_1$, 
as well as $\wit{\mathfrak{age}}_1$ and $\wit{\mathfrak{alt}}_1$ 
because they do not contain the time-translation generator $X_{-1}$. 

After these preparations, we can now formulate the questions studied in this 
paper. For notational simplicity, we restrict ourselves to $d=1$ space
dimension and look for a semilinear and non-derivative extension of the
free ``Schr\"odinger equation'' of the form 
\BEQ \label{gl:Sop}
\hat{\cal S}\Psi := \left( 2\partial_{\zeta}\partial_t-\partial^2_r\right)\Psi 
= F(\zeta,t,r,\Psi ,\Psi^*)
\EEQ
which are invariant under one of the subalgebras of 
$(\mathfrak{conf}_3)_{\mathbb{C}}$ as sketched in figure~\ref{Bild1}bcd. 
We shall initially 
take $\Psi$ to be a complex-valued function as is appropriate for the
physical context of eq.~(\ref{eq:NLSE}) and shall return later to the
question whether non-trivial symmetries involving only real-valued functions 
$\Psi$ are possible, as would 
appear more natural for equation (\ref{modelA}). Our main tool will be
the construction of new differential-operator representations of the
algebras defined above. In section~2, we shall relax the condition of
having skew-hermitian representations of the Schr\"odinger group 
and of the other 
algebras of figure~\ref{Bild1}. More general representations will be 
constructed and we shall then find the most
general semi-linear equation of the form (\ref{gl:Sop}) invariant under these. 
On the other hand, non-linear equations of the forms 
(\ref{eq:NLSE},\ref{modelA},\ref{gl:Sop}) imply the existence of a
coupling constant with the nonlinearities, which in existing studies is 
tacitly admitted to be dimensionless. This hidden assumption leads to rather 
restricted form of admissible potentials. Hence for a given form of the
potential, Schr\"odinger-invariance should only hold for one special value 
of the spatial dimensionality $d$, which is in disagreement with the existing 
numerical and analytical results on ageing quoted above. 
We shall inquire into the consequences of 
treating dimensionful coupling constants, 
which consequently will transform under scaling and special
transformations. In section~3, we contruct the representations with a 
dimensionful coupling constant and in section~4 we find the
nonlinear Schr\"odinger equations invariant under these new representations. 
Section~5 presents our conclusions.

\section{Non skew-hermitian representations}

\subsection{General remarks}

We recall first the basic method used in finding non-linear equations with a
given symmetry group, as outlined e.g. in \cite{Boye76}. 
Consider first the linear equation
\BEQ
\hat{S} \Phi(t,r)=0 \label{eq:line}
\EEQ
and let $G$ be an one-dimensional Lie group which acts on the equation 
(\ref{eq:line}) such that solutions are transformed into solutions. 
In this paper we are going to consider a projective action of the
elements $g\in G$ on the solutions $\Phi(t,r)$ such that
\BEQ
[T(g)\Phi ](t,r)=\mu_g(t,r)\Phi(t',r'). \label{eq:action}
\EEQ   
where $(t',r')=g(t,r)$. For $G$ to be symmetry group one requires
\BEQ
\hat S[T(g)\Phi ](r,t)=0 \label{eq:simc}
\EEQ
for all $\Phi $ satisfying (\ref{eq:line}). It is convenient to consider
instead of $G$ its Lie algebra $\mathfrak{g}$ and to expand
\BEA
T(g)  &=& 1+\eps X+ \ldots\nonumber \\
X\Phi &=& \left(a(t,r)\partial_t+b(t,r)\partial_r+c(t,r)\right)\Phi 
\label{eq:expt}
\EEA
Now, equation (\ref{eq:simc}) implies the operator equation
\BEQ
[\hat{S}, X]=\lambda (t,r)\hat{S}\label{eq:ginl}
\EEQ
In the same way, for the nonlinear equation 
\BEQ 
\hat S \Phi(t,r)=F(t,r,\Phi ,\Phi^*) \label{eq:nline}
\EEQ
one arrives at the following conditions
\BEA
\hat{S}[T(g)\Phi ](t,r) &=& 
F\left(t,r,[T(g)\Phi ](t,r),[T(g)\Phi ]^*(t,r)\right) 
\nonumber \\
\hat{S}(X\Phi ) &=& (X\Phi)\partial_{\Phi }F+(X\Phi )^*\partial_{\Phi^*}F 
\label{eq:pinl}
\EEA
Combining (\ref{eq:ginl},\ref{eq:pinl}) with the explicit form 
(\ref{eq:expt}) of $X$ gives \cite{Boye76}
\BEQ \label{eq:repot}
\left[a(t,r)\partial_t+b(t,r)\partial_r-c(t,r)\Phi \partial_{\Phi}
-c^*(t,r)\Phi^*\partial_{\Phi^*}+(c(t,r)+\lambda(t,r))\right]
F(t,r,\Phi,\Phi^*)=0 
\EEQ
This linear equation can be solved with standard techniques. For Lie algebras
with $\dim \mathfrak{g}>1$, one has one such equation 
for each independent generator. 
In the same way, further independent variables can be included 
straightforwardly. Evidently, the symmetry algebra of the quasi-linear 
equations considered here 
must be a subalgebra of the symmetry algebra of the linear part. 

Throughout, we shall work with the linear Schr\"odinger operator
\BEQ \label{eq:cassch}
\hat{S} := 2M_0 X_{-1} - Y_{-\frac{1}{2}}^2
\EEQ 
which satisfies 
\BEA
{}[\hat S,M_0] &=&[\hat S,Y_{-\frac{1}{2}}] \:=\: [\hat S,Y_{1\over 2}] \:=\:
[\hat{S},X_{-1}] \:=\:  [\hat{S},N] \:=\:  0 \nonumber \\
{}[\hat S,X_0]&=& -\hat S \;\; , \;\; 
[\hat S,X_1] \:=\: -2t\hat S  -2 \left( x -\frac{1}{2}\right) M_0
\label{eq:comsh} \\
{}[\hat{S},V_+] &=& -4r \hat{S} - 4\left(x-\frac{1}{2}\right)Y_{-\frac{1}{2}}
\nonumber 
\EEA
{}from which we can read off the eigenvalues $\lambda(t,r)$. We also see 
that invariance under the algebras $\mathfrak{sch}_1$, $\mathfrak{age}_1$ or
$\mathfrak{alt}_1$  (or the parabolic extensions) holds on the space
of solutions of $\hat{S}\Phi=0$, provided $x=1/2$. 

\subsection{Representation with a generalized scaling behaviour}

For the free Schr\"odinger equation $\hat{S}\Phi=0$, invariance under
$\mathfrak{sch}_1$ implies that the scaling dimension of the solution
$\Phi$ is $x=\frac{1}{2}$. In order to relax this constraint while
conserving the symmetry, we follow \cite{Pico04} and modify the 
generator $X_1$ in (\ref{eq:zetaGen}) by adding a term $-k t$, 
where $k$ is some constant. Then we should further add a further 
generator $Z : = -k Z_0$, where $Z_0$ is central. From now on, we shall work
with a variable `mass' and go over to the conjugate variable $\zeta$. 
We require that the commutators which are not in the centre are unchanged 
and find that with respect to 
eqs.~(\ref{eq:zetaGen},\ref{eq:zetaGen2}) only the following generators are
modified
\BEA
X_1 &=& -t^2\partial_t-tr\partial_r-{1\over 2}r^2\partial_{\zeta }-(x+k)t \\
V_+&=& -2tr\partial_t-2\zeta r\partial_{\zeta}
       -(r^2+2\zeta t)\partial_r-2(x+k)r \nonumber \\
Z &=& -kZ_0 \nonumber
\EEA
The only commutators which are modified are the central ones 
\BEA
{}[X_1,X_{-1}] &=& 2X_0 + k Z_0 \nonumber \\
{}[V_+,Y_{-{1\over 2}}] &=& 4X_0-2N+2kZ_0 \label{eq:tdc}
\EEA
and we see that $Z_0$ can be absorbed into a redefinition of the 
generators $X_0$ or $D$, as expected on general grounds \cite{Kac87}. 
However, the presence of the parameter $k$ is important for the
determination of invariant non-linear equations, as we shall see now. 

Consider the non-linear Schr\"odinger equation (\ref{gl:Sop})
\BEQ
{(2\partial_{\zeta }\partial_t-\partial^2_r)\Psi (\zeta ,t,r) 
= {\tilde F}(\zeta,t,r,\Psi ,\Psi^*)}
\EEQ
Eq.~(\ref{eq:repot}) then gives the following conditions on $\tilde{F}$, 
for each of the generators: 
\BEA
Y_{-{1\over 2}} &:& \partial_r{\tilde F} = 0 \label{eq:sp} \\
X_{-1}          &:& \partial_t{\tilde F} = 0 \label{eq:tm} \\
M_0             &:& \partial_{\zeta}{\tilde F} = 0 \label {eq:mas}\\
Y_{1\over 2}    &:& (t\partial_r+r\partial_{\zeta}){\tilde F}= 0 
\label{eq:gl} \\
X_0             &:& \left[t\partial_t +{1\over 2}r\partial_r
                   -{x\over 2}(\Psi \partial_{\Psi }+\Psi^*\partial_{\Psi^*})
                   +\left(\frac{x}{2}+1\right)\right]{\tilde F} = 0 
\label{eq:dl} \\
D               &:& \left[ t\partial_t +r\partial_r +\zeta \partial_{\zeta}
                   -x(\Psi \partial_{\Psi }+\Psi^*\partial_{\Psi^*})
                    +(x+2)\right]{\tilde F} = 0 
\label{eq:compdl} \\
N               &:& (t\partial_t -\zeta\partial_{\zeta}){\tilde F} = 0 
\label{eq:ce} \\
X_1             &:& \left[ t^2\partial_t+tr\partial_r
                    +{r^2\over 2}\partial_{\zeta}
                    -(x+k)t(\Psi\partial_{\Psi}+\Psi^*\partial_{\Psi^*})
                    +(x+k+2)t\right]{\tilde F} = 0
\label{eq:cf} \\
V_+             &:& \left[2tr\partial_t+(2t\zeta+r^2)\partial_r
                   +2r\zeta\partial_{\zeta}
                   -2r(x+k)(\Psi \partial_{\Psi}+\Psi^*\partial_{\Psi^*})
                   +2r(x+k+2)\right]{\tilde F} = 0 
\label{eq:ph} 
\EEA

We point out that already the invariance of the {\em linear}
Schr\"odinger equation under $X_1$ requires $x+k={1\over 2}$. Consequently,
the scaling dimension of $\Psi$ which has to be fixed for an invariance is
given by $x+k$ and {\em not} by $x$ alone. This is a
necessary condition for the invariance of the nonlinear equation. 

Our results are as follows.
\begin{enumerate}
\item \underline{The algebra ${\mathfrak{sch}}_1$.} 
We have to solve the system
eqs.~(\ref{eq:sp},\ref{eq:tm},\ref{eq:mas},\ref{eq:gl},\ref{eq:dl},\ref{eq:cf}).
A solution only exists in the well-known case $k=0$, hence $x=\frac{1}{2}$. 
This in indeed quite analogous to the equation (\ref{eq:NLSE}) with a fixed mass where the same power-law for the potential was found, 
see e.g. \cite{Boye76,Baru80,Fush93,Ride93}. One has
\BEQ \label{2:sch}
{\tilde F}_{\mathfrak{sch}_1} =  {\tilde F}_{\mathfrak{sch}_1}(\Psi,\Psi^*) 
= \Psi^5 f\left({\Psi\over {\Psi}^*}\right).
\EEQ
where $f$ denotes an arbitray differentiable function. 
Furthermore, adding the generator $N$ by taking eq.~(\ref{eq:ce}) into account 
does not give anything new. Hence the result for the parabolic subalgebra
$\wit{\mathfrak{sch}}_1$ is the same as before, viz. 
$\tilde{F}_{\wit{\mathfrak{sch}}_1}=\tilde{F}_{\mathfrak{sch}_1}$.\footnote{In
view of the embedding $\mathfrak{sch}_d \subset (\mathfrak{conf}_{d+2})_\mathbb{C}$ one might also consider the semi-linear
conformally invariant Klein-Gordon equation $\Box \phi = F(\phi) = \phi^{(n+2)/(n-2)}$ in $n=d+2$ dimensions, e.g. \cite{Fush93,Shat00}, which reproduces the form $F\sim \phi^5$ for the case $d=1$.} 

In consequence, no new form of an invariant nonlinear equation is found. 
\item \underline{The algebra ${\mathfrak{age}}_1$.} We now have to solve
the sytems eqs.~(\ref{eq:sp}, \ref{eq:mas}, \ref{eq:gl}, \ref{eq:dl}, 
\ref{eq:cf}). There is no further constraint on $k$ and the solution is,
using again $x+k=1/2$ 
\BEQ \label{2:age}
{\tilde F}_{\mathfrak{age}_1} = {\tilde F}_{\mathfrak{age}_1}(\Psi,\Psi^*;t)  
= t^{-4k/(2k+1)}{\Psi}^{(2k+5)/(2k+1)} 
f\left({\Psi\over {\Psi}^*}\right).
\EEQ
and we see that there also appears an explicit dependence on time
(only the special case $k=0$ was considered in \cite{Boye76}). 
On the other hand, if we go to the parabolic subalgebra 
$\wit{\mathfrak{age}}_1$ by adding the 
generator $N$ we must have $k=0$ and we find 
${\tilde F}_{\wit{\mathfrak{age}}_1} = {\tilde F}_{\mathfrak{sch}_1}$, 
that is the same result as in the Schr\"odinger case.
\item \underline{The algebra ${\mathfrak{alt}}_1$.}
Here we must solve the equations (\ref{eq:sp}, \ref{eq:mas}, \ref{eq:gl}, 
\ref{eq:compdl}, \ref{eq:cf}, \ref{eq:ph}). There are two distinct solutions.
The first one exists for all values of $k$ and reads 
\BEQ \label{2:alt1}
{\tilde F}_{\mathfrak{alt}_1}^{(1)} 
= {\tilde F}_{\mathfrak{alt}_1}^{(1)}(\Psi,\Psi^*;t) 
= t^{-2}\Psi f\left({\Psi\over {\Psi}^*}\right).
\EEQ
However, for $k=0$ there is a second solution 
\BEQ \label{2:alt2}
{\tilde F}_{\mathfrak{alt}_1}^{(2)} 
= {\tilde F}_{\mathfrak{alt}_1}^{(2)}(\Psi,\Psi^*;t) 
= {\tilde F}_{\mathfrak{sch}_1}(\Psi,\Psi^*)
\EEQ
Finally, if we go over to $\wit{\mathfrak{alt}}_1$ by adding the
generator $N$, a solution exists only for $k=0$, hence 
${\tilde F}_{\wit{\mathfrak{alt}}_1} = {\tilde F}_{\mathfrak{sch}_1}$. 
\end{enumerate}
A few observations are in order. First, we see that we can trade in some cases
the dependence of the potential $\tilde{F}$ on $\Psi$ in favour of an
explicit dependence on $t$ such that the total dimension of $\tilde{F}$ is
unchanged. Second, we find that this freedom does not exist for the 
parabolic subalgebras but only for the almost-parabolic subalgebras
$\mathfrak{age}_1$ and $\mathfrak{alt}_1$ which do not contain the
time-translations. Third, and in contrast to the 
well-known results where the `masses'
$m$ (or $\cal M$) are kept fixed \cite{Boye76,Fush93,Ride93}, 
the dependence on the phase $\Psi/\Psi^*$ is not determined in these 
representations. In particular, if we take $f={\rm cste.}$, we even have
the possibility to consider a Schr\"odinger equation with a {\em real}
solution $\Psi$ and a non-trivial symmetry, which is impossible if the
masses are kept fixed. 

While we are not aware of immediate physical applications of the equations
found here, these examples suggest that it may be useful to consider the
presence of other dimensionful quantities, such as coupling constants, in the
nonlinear equations. We shall take this up in the next section. 

\section{Representations with an additional dimensionful coupling}

In our search for more general quasilinear Schr\"odinger equations with
a non-trivial symmetry we now consider explictly a dimensionful coupling
constant $g$ in the nonlinear term. In what follows, we do
consider that $g$ also changes under the local scale-transformations. 
Hence the dilatation generator is taken to be of the form
\BEQ
X_0 =-t\partial_t-{1\over 2}r\partial_r-yg\partial_g-{x\over 2} \label{eq:SC}
\EEQ
where $y$ is the scaling dimension of the coupling $g$. In addition, we also
assume throughout that the generator of space translations 
$Y_{-{1\over 2}}=-\partial_r$ is unchanged. 

In this section we construct the remaining generators. Their explicit
form will be used in section~4 to find the invariant semilinear equations. 

\newpage\typeout{ *** Hier ist ein Seitenvorschub ! ***}
\subsection{Subalgebras ${\mathfrak{age}}_1$ and $\wit{\mathfrak{age}}_1$}

Starting form eqs.~(\ref{eq:zetaGen},\ref{eq:zetaGen2}), we look for 
extensions of the usual generators and write 
\BEA
M_0          &=&-\partial_{\zeta}-L(t,r,\zeta ,g)\partial_g
\nonumber \\
Y_{1\over 2} &=& -t\partial_r-r\partial_{\zeta }-Q(t,r,\zeta ,g)\partial_g 
\nonumber \\
X_1   &=& -t^2\partial_t-tr\partial_r
          -{1\over 2}r^2\partial_{\zeta }-P(t,r,\zeta ,g)\partial_g-xt  
\label{eq:CONF} \\
N     &=& -t\partial_t+\zeta \partial_{\zeta }
          -K(t,r,\zeta ,g)\partial_g 
\nonumber
\EEA
The unknown functions $L,Q,P,K$ are determined from the following 
two requirements: 
\begin{enumerate}
\item the commutators of the standard realizations of $\mathfrak{age}_1$ and
$\wit{\mathfrak{age}}_1$ are assumed to remain valid. 
\item invariance of the linear Schr\"odinger
equation under the new representations.
\end{enumerate}

First, we consider the  almost-parabolic subalgebra $\mathfrak{age}_1$. 
Imposing the commutator relations, our results are as follows. 
The commutators
\BEQ
[X_0,X_1]=-X_1\;\;,\;\;
[X_0,Y_{1\over 2}]=-{1\over 2}Y_{1\over 2} \;\;,\;\;
[X_0,M_0]=0
\EEQ 
give
\BEQ
P = p_0(\zeta )t^{y+1} p(u,v)\;\; , \;\;  
Q = q_0(\zeta )t^y q(u,v)\;\; , \;\; 
L = l_0(\zeta )t^{y-1} l(u,v) 
\EEQ
where $u={r^2/t}$ and $v={t^y/g}$. Next we use
\BEQ
[X_1,Y_{-{1\over 2}}]=Y_{1\over 2} \;\;,\;\;
[Y_{1\over 2},Y_{-{1\over 2}}]=M_0 \;\;,\;\;
[Y_{-{1\over 2}},M_0]=0 \;\;,\;\;
[Y_{1\over 2},M_0]=0
\EEQ
and obtain
\BEQ
q(u,v) =2u^{1\over 2}\partial_up(u,v) \;\;,\;\;
l(u,v) =2u^{1\over 2}\partial_u q(u,v) \;\; , \;\; 
\partial_u l(u,v) = 0 \;\; , \;\; 
p_0(\zeta ) = q_0(\zeta )=l_0(\zeta )
\EEQ 
Hence $l=l(v)$ and 
\BEQ 
q(u,v) = u^{1/2}l(v)+n(v) \;\;,\;\;
p(u,v) = {u\over 2}l(v)+u^{1/2}n(v)+m(v) 
\EEQ
where $m(v), n(v)$ are two functions of $v$ to be determined.
The last remaining commutators 
\BEQ           
[X_1,Y_{1\over 2}]=0\;\;,\;\;[X_1,M_0]=0
\EEQ
lead to the following system for the yet unknown functions
$l(v), n(v), m(v),p_0(\zeta)$, where the prime denotes the derivative
with respect to the argument 
\BEA
p_0(\zeta )y\left(l(v)+vl'(v)\right)+p_0^2(\zeta)v^2 \left(l(v)m'(v)-m(v)l'(v)\right) 
-p'_0(\zeta)m(v)&=&0 \label {eq:syu}\\ 
(2y-1)n(v)+2yvn'(v)+2p_0(\zeta)v^2\left(n(v)m'(v)-m(v)n'(v)\right) &=& 0 
\label {eq:syd}\\ 
p_0'(\zeta)n(v)-p_0^2(\zeta)v^2\left(l(v)n'(v)-n(v)l'(v)\right) &=& 0 
\label {eq:syt} 
\EEA
Performing a separation of the variables $v$ and $\zeta$ in eq.~(\ref{eq:syt}),
we see that two cases must be distinguished, depending on whether $p_0'(\zeta)$
vanishes or not.

\noindent \underline{{\bf Case a)} 
The {\it NMG-representation} with a non-modified mass generator $M_0$.}
In this case, one has $p_0(\zeta )=p_{01}=\mbox{\rm cste}$. 
We have $n(v)=l(v)=0,m(v) \ne 0$ and consequently:
\BEQ
L=0\;\;,\;\; Q=0\;\;, \;\; P=p_{01}t^{y+1}m(v) \label{eq:solu}
\EEQ
and the function $m(v)$ remains arbitrary. 

\noindent \underline{{\bf Case b)} The {\it MMG-representation} with a modified
mass generator $M_0$.} In this case, we have 
$p_0(\zeta)=-\frac{2y}{l_0\zeta}$ where $l_0$ is a constant.
We find the solutions $l(v)=l_0v^{-1}$, $n(v)=n_0v^{(1-2y)/(2y)}$ and
$m(v)=cn(v)$ and set $h_0:={n_0/l_0}$. Consequently
\BEA
L &=&-\frac{2y}{\zeta }g\;\;, \;\; 
Q=-\frac{2y}{\zeta}\left(rg+h_0g^{(2y-1)/(2y)}\right)\nonumber \\
P &=&-\frac{2y}{\zeta}\left[{r^2g\over 2}+h_0trg^{(2y-1)/(2y)}
+ch_0t^{3/2}g^{(2y-1)/(2y)}\right]\label{eq:sold}
\EEA

We have hence established the existence of two distinct representations
of the algebra $\mathfrak{age}_1$. Explicitly, we always have 
$Y_{-1/2}=-\partial_r$ and further in the case a) of the NMG-representation 
\BEA
M_0 &=& -\partial_{\zeta} \nonumber \\
Y_{1\over 2} &=& -t\partial_r-r\partial_{\zeta } \nonumber \\
X_0 &=& -t\partial_t-{1\over 2}r\partial_r-yg\partial_g-{x\over 2} \nonumber \\
X_1 &=& -t^2\partial_t-tr\partial_r -{1\over 2}r^2\partial_{\zeta}
        -p_{01}t^{y+1}m\left({t^y/g}\right)\partial_g-xt
\label{eq:ageun} 
\EEA
The NMG-representation of $\mathfrak{age}_1$ is parametrized by the three 
constants $x,y,p_{01}$ and an arbitrary function $m(v)$. 
In the case b) of the MMG-representation we have
\BEA
M_0 &=& -\partial_{\zeta }+\frac{2y}{\zeta }g\partial_g 
\nonumber \\
Y_{1\over 2} &=& -t\partial_r-r\partial_{\zeta}
+\frac{2y}{\zeta}
\left(rg+h_0g^{(2y-1)/(2y)}\right)\partial_g\nonumber \\
X_0 &=& -t\partial_t-{1\over 2}r\partial_r-yg\partial_g-{x\over 2} \nonumber \\
X_1 &=& -t^2\partial_t-tr\partial_r-{1\over 2}r^2\partial_{\zeta } \nonumber \\
&& +\frac{2y}{\zeta}\left({r^2g\over 2}+h_0trg^{(2y-1)/(2y)}
+ch_0t^{3/2}g^{(2y-1)/(2y)}\right)\partial_g-xt \label{eq:agdeux}
\EEA
The MMG-representation of $\mathfrak{age}_1$ is parametrized by the four 
constants $x,y,h_0,c$.

We now look for the most general representation of the parabolic 
subalgebra $\wit{\mathfrak{age}}_1$. From the commutators
\BEQ
[X_0,N]=0\;\;,\;\;[Y_{-{1\over 2}},N]=0
\EEQ
we obtain
\BEQ
K=k_0(\zeta )t^yk(u,v)\;\;,\;\;\partial_uk(u,v)=0 \label{eq:detk}
\EEQ
hence $k(u,v)=k(v)$. Next, from the commutators
\BEQ
[X_1,N]=X_1\;\;,\;\;[Y_{1\over 2},N]=Y_{1\over 2}\;\;,\;\;[M_0,N]=M_0
\EEQ
we obtain that the functions $l(v)$, $n(v)$, $m(v)$, $k(v)$, $p_0(\zeta)$ and
$k_0(\zeta)$  must satisfy the system:
\BEA
{yk_0(\zeta)\left(k(v)+vk'(v)\right)-yp_0(\zeta)\left(m(v)+vm'(v)\right)
+\zeta p_0'(\zeta)m(v)} & &  \nonumber \\
-k_0(\zeta)p_0(\zeta)v^2\left(m(v)k'(v)-m'(v)k(v)\right) &=& 0 
\label{eq:suq}\\
{2yp_0(\zeta)\left(n(v)+vn'(v)\right)
-p_0(\zeta)n(v)-2\zeta p_0'(\zeta)n(v)} & &  \nonumber \\
+ 2k_0(\zeta)p_0(\zeta)v^2(n(v)k'(v)-n'(v)k(v)) &=& 0 
\label{eq:suc}\\ 
{yp_0(\zeta)(l(v)+vl'(v))-(p_0(\zeta)+\zeta p_0'(\zeta))l(v)} & & 
\nonumber \\
+k_0(\zeta)p_0(\zeta)v^2(l(v)k'(v)-l'(v)k(v))-k_0'(\zeta )k(v) &=& 0 
\label{eq:sus} 
\EEA
Our results from above for the functions $l(v),n(v),m(v)$ found for 
the algebra $\mathfrak{age}_1$ lead to:

\noindent 
\underline{{\bf Case a)} NMG-representation.} \\
The nontrivial equations are (\ref{eq:suq}) and (\ref{eq:sus}).
The last leads to $ k_0'(\zeta )k(v) = 0 $ which is satisfied
if $k_0(\zeta )=k_0={\rm cste.}$ (The other case $k(v)=0$ leads
to $K=0, m(v)=m_0v^{-1}$, $m_0={\rm cste.}$, which one can equivalently 
obtain by putting $k_0=0$ in (\ref{eq:suq}).)

We find, besides $Y_{-1/2} =-\partial_r$ and $M_0 = -\partial_{\zeta}$ for the
NMG-representation of $\wit{\mathfrak{age}}_1$
\BEA
Y_{1/2} &=& -t\partial_r-r\partial_{\zeta } \nonumber \\
X_0 &=& -t\partial_t-{1\over 2}r\partial_r-yg\partial_g-{x\over 2} 
\nonumber \\
X_1 &=& -t^2\partial_t-tr\partial_r-{1\over 2}r^2\partial_{\zeta}
-p_{01}t^{y+1}m\left({t^y\over g}\right)\partial_g-xt\nonumber \\
N &=& -t\partial_t+\zeta\partial_{\zeta}-k_0 t^y 
k\left({t^y\over g}\right)\partial_g \label{eq:nageun} 
\EEA 
This representation is characterized by the constants $x,y,p_{01},k_0$ 
and the functions $m(v),k(v)$ which must satisfy the condition
\BEQ
yk_0\frac{\D}{\D v}(vk(v))-yp_{01}\frac{\D}{\D v}(vm(v))- 
k_0p_{01}v^2k(v)^2\frac{\D}{\D v}\left(\frac{m(v)}{k(v)}\right)=0 
\label{eq:arbc}
\EEQ

\noindent 
\underline{{\bf Case b)} MMG-representation.} \\
In this case the equations (\ref{eq:suq}) and (\ref{eq:suc}) lead to
\BEQ
yk_0(\zeta )(k(v)+vk'(v))+p_0(\zeta )m(v)=0 \label{eq:agpm} 
\EEQ
Two cases must be considered:

{\bf 1.} $k_0(\zeta )=k_0={\rm cste.}$. 
This leads to $k(v)=\varkappa v^{-1}$, $n(v)=m(v)=0$ and the first 
MMG-realization of the algebra $\wit{\mathfrak{age}}_1$ is
\BEA
M_0 &=& -\partial_{\zeta }+\frac{2y}{\zeta }g\partial_g 
\nonumber \\
Y_{1\over 2} &=& -t\partial_r-r\partial_{\zeta}
+\frac{2y}{\zeta}\left(rg\right)\partial_g\nonumber \\
X_0 &=& -t\partial_t-{1\over 2}r\partial_r-yg\partial_g-{x\over 2} \nonumber \\
X_1 &=& -t^2\partial_t-tr\partial_r-{1\over 2}r^2\partial_{\zeta } 
        +\frac{2y}{\zeta}\left({r^2g\over 2}\right)\partial_g-xt \nonumber \\
N &=& -t\partial t+\zeta \partial_{\zeta }-k_0g\partial_g \label{eq:nagdeux}
\EEA 

{\bf 2.} $k_0'(\zeta )\ne 0$.   
We divide the equation (\ref{eq:agpm}) by $(k_0(\zeta )p_0(\zeta))$ 
(since $k_0(\zeta)\ne 0,p_0(\zeta )\ne 0$),
denote $\mathcal{K}=1/k_0(\zeta )$ and take the derivative with 
respect to $\zeta $ for obtaining
\BEQ
{\mathcal{K}}'(\zeta  )=l_0{(k(v)+vk'(v))\over m(v)}=:C_0 = \mbox{\rm cste.}
\EEQ
with solutions
\BEQ
k_0(\zeta )=\frac{1}{C_0\zeta }\;\;,\;\;k(v)=2y\frac{C_0}{l_0}m(v)=\varkappa v^{(1-2y)/(2y)}\;\;,\;\;
\varkappa=2yC_0ch_0 \label{eq:kcent}
\EEQ
The second MMG-realization of the algebra $\wit{\mathfrak{age}}_1$ is
\BEA
M_0 &=& -\partial_{\zeta }+\frac{2y}{\zeta }g\partial_g 
\nonumber \\
Y_{1\over 2} &=& -t\partial_r-r\partial_{\zeta}
+\frac{2y}{\zeta}\left(rg+h_0g^{(2y-1)/(2y)}\right)\partial_g\nonumber \\
X_0 &=& -t\partial_t-{1\over 2}r\partial_r-yg\partial_g-{x\over 2} \nonumber \\
X_1 &=& -t^2\partial_t-tr\partial_r-{1\over 2}r^2\partial_{\zeta } \nonumber \\
&& +\frac{2y}{\zeta}\left({r^2g\over 2}+h_0trg^{(2y-1)/(2y)}
+ch_0t^{3/2}g^{(2y-1)/(2y)}\right)\partial_g-xt \nonumber \\
N &=& -t\partial t+\zeta \partial_{\zeta }-\frac{ch_0}{\zeta }t^{1\over 2}g^{(2y-1)/(2y)}\partial_g \label{eq:cexnagdeux}
\EEA

Having exhausted the constraints from the commutation relation, we now require
that the solution space of the {\em linear} Schr\"odinger equation is invariant
under the action of these representations. This means  
\BEQ
[\hat S,X_i]=\lambda_i\hat S \label{eq:schinv},
\EEQ
for each of the generators, where $\hat{S}$ is in the 
representation-independent form (\ref{eq:cassch}). We must distinguish 
the two cases
\begin{enumerate}
\item a fixed scaling dimension $x=1/2$ for the wave function $\Psi$. 
\item wave functions with arbitrary scaling dimensions.
\end{enumerate}
First, for the NMG-realization (\ref{eq:ageun}) of the 
algebra $\mathfrak{age}_1$ we have 
\BEA
{}[\hat{S},X_0] &=& -\hat{S} \nonumber \\
{}[\hat{S},M_0] &=& [\hat{S},Y_{-{1\over 2}}] \:=\: 
{}[\hat S,Y_{1\over 2}]\:=\: 0 \nonumber \\
{}[\hat S,X_1] &=&-2t\hat S -(1-2x)M_0+M_0\hat Q \label{eq:cfshin} 
\EEA
where 
\BEQ
\hat{Q} := \left[2p_{01}t^y\left((y+1)m(v)+yvm'(v)\right)\right]\partial_g.
\EEQ
In order to satisfy the condition (\ref{eq:schinv}) one must have
\BEQ
\left((1-2x)M_0-M_0\hat Q\right)\Psi =0 \label{eq:oper}
\EEQ
In general, we want to satisfy the (\ref{eq:oper}) on the operator level
and shall relax this to a condition on the functions in the representation
space only if a non-trivial solution cannot be found otherwise. 

We begin with the case $x={1\over 2}$ where we must have
$\hat Q\Psi =(2p_{01}t^y((y+1)m(v)+yvm'(v)))\partial_g\Psi =0$. 
This leads to
\BEQ
m(v)=m_0v^{-{(y+1)/y}} \label{eq:ncpo}
\EEQ
The final NMG-realization of $\mathfrak{age}_1$ is in the case at hand
\BEA
Y_{-{1\over 2}} &=& -\partial_r \nonumber \\
M_0 &=& -\partial_{\zeta} \nonumber \\
Y_{1\over 2} &=& -t\partial_r-r\partial_{\zeta } \nonumber \\
X_0 &=& -t\partial_t-{1\over 2}r\partial_r-yg\partial_g-{1\over 4} 
\nonumber \\
X_1 &=& -t^2\partial_t-tr\partial_r-{1\over 2}r^2\partial_{\zeta}
        -p_{01}m_0g^{(y+1)/y}\partial_g-{1\over 2}t
\label{eq:rxageun} 
\EEA
and the invariant linear Schr\"odinger equation is simply
$\left( 2\partial_{\zeta}\partial_{t}-\partial_{r}^2\right)\Psi=0$.

In the other case of arbitrary scaling dimension $x\ne 1/2$ we have 
\BEQ
\hat Q\Psi =(1-2x)\Psi \label{eq:cgox}
\EEQ
to be satisfied. Here, because of the arbitrary value of $m(v)$ the condition 
(\ref{eq:cgox}) is written in integral form 
(using $\Psi(t,r,\zeta,g)\longrightarrow \Psi(t,r,\zeta,v)$ where $v=t^y/g$):
\BEQ \label{age0condition}
\Psi(t,r,\zeta,v)=\Psi_0(t,r,\zeta){1-2x\over 2p_{01}}\int {\D v\over v^2((y+1)m(v)+yvm'(v))}
\EEQ
and the NMG-realisation of the algebra $\mathfrak{age}_1$ is 
given by eq.~(\ref{eq:ageun}).

We remember now, that {\it ``almost-parabolic subalgebras''} were defined through parabolic
ones and we anticipate a result of the parabolic subalgebra
$\wit{\mathfrak{age}}_1$. In the NMG-representation, we shall show below
that invariance of the  linear Schr\"odinger equation leads to $k(v)=v^{-1}$.
Hence, because of (\ref{eq:arbc}) we have $m(v)=v^{-1}$ and 
\BEQ
\hat{Q}= 2p_{01}g\partial_g \label{eq:ansatz}
\EEQ
for $x\ne {1\over 2}$ and $Q=0\;\;(p_{01}=0)$ in the
case of canonical scaling dimension $x={1\over 2}$.

Invariance of the linear Schr\"odinger equation implies the following condition on the wave function 
\BEQ \label{eq:restr}
\hat Q \Psi(t,r,\zeta ,g) = 
2p_{01}g\partial_g\Psi(t,r,\zeta ,g) = (1-2x)\Psi(t,r,\zeta ,g) 
\EEQ
which means $\Psi(t,r,\zeta, g) = g^{(1-2x)/(2p_{01})}\psi(t,r,\zeta)$
and the dependence of $\Psi$ on $g$ is determined. 

There is another possibility, namely $M_0\Psi=0$.
In this case the wave functions do not depend on $\zeta$.
So we have seen that instead of a simple Lie symmetry, we 
rather have a so-called conditional symmetry as introduced in 
\cite{Fush93,Levi89}. 

Finally, the NMG-representation of the algebra  $\mathfrak{age}_1$ for
arbitrary scaling dimensions, subject to the auxiliary condition 
(\ref{eq:restr}) is
\BEA
Y_{-{1\over 2}} &=& -\partial_r \nonumber \\
M_0 &=& -\partial_{\zeta} \nonumber \\
Y_{1\over 2} &=& -t\partial_r-r\partial_{\zeta } \nonumber \\
X_0 &=& -t\partial_t-{1\over 2}r\partial_r-yg\partial_g-{x\over 2} 
\nonumber \\
X_1 &=& -t^2\partial_t-tr\partial_r-{1\over 2}r^2\partial_{\zeta}
        -p_{01}tg\partial_g-xt
\label{eq:rageun} 
\EEA

Second, we consider the MMG-realization of (\ref{eq:agdeux}) of the 
algebra $\mathfrak{age}_1$. From the conditions (\ref{eq:schinv}) 
we have $h_0=0$ and hence
\BEA
{}[\hat S,X_0] &=&-\hat S \nonumber \\
{}[\hat S,M_0] &=& [\hat S,Y_{-{1\over 2}}] \:=\: 
{}[\hat S,Y_{1\over 2}] \:=\: 0 \nonumber \\
{}[\hat S,X_1] &=& -2t\hat S -(1-2x)M_0.
\EEA
which are satisfied automatically in the case $x={1\over 2}$.

{\bf Remark 1:} The subalgebra of $\mathfrak{age}_1$ obtained when 
leaving out the generator $X_1$ leaves the linear Schr\"odinger 
equation invariant for arbitrary $h_0$.
  
In the case of an arbitrary scaling dimension we must have  
\BEQ \label{eq:nerest}
M_0\Psi(t,r,\zeta,g)=\left(\partial_{\zeta}-\frac{2y}{\zeta}g\partial_g\right)\Psi(t,r,\zeta ,g)=0 
\EEQ
which implies $\Psi(t,r,\zeta,g)=\Psi(t,r,\zeta g^{1/2y})$
such that the dependence on $\zeta$ and $g$ merely enters through the
scaling variable $u :=g^{1/2y}\zeta $. Again, we merely find
a conditional symmetry. 

Our final result for the MMG-representation of $\mathfrak{age}_1$ reads, for
any value of $x$ 
\BEA
Y_{-{1\over 2}} &=& -\partial_r \nonumber \\
M_0 &=& -\partial_{\zeta }+\frac{2y}{\zeta}g\partial_g 
\nonumber \\
Y_{1\over 2} &=& -t\partial_r-r\partial_{\zeta}+\frac{2y}{\zeta}rg\partial_g\nonumber \\
X_0 &=& -t\partial_t-{1\over 2}r\partial_r-yg\partial_g-{x\over 2} \nonumber \\
X_1 &=& -t^2\partial_t-tr\partial_r-{1\over 2}r^2\partial_{\zeta }
+\frac{y}{\zeta}r^2g\partial_g-xt \label{eq:rnagdeux} 
\EEA

We finish by extending our results to the parabolic 
subalgebra $\wit{\mathfrak{age}_1}$.  For the NMG-representation 
we must also satisfy the condition
\BEQ
[\hat S,N]=yt^{y-1}k_0(k(v)+vk'(v))=0
\EEQ
which gives $k(v)=v^{-1}$ and hence, from (\ref{eq:arbc}) we find $m(v)=v^{-1}$
for the general case and $m(v)=0$ in the special case $x={1\over 2}$. 
The generators of the NMG-representation of $\wit{\mathfrak{age}}_1$ read
\BEA
Y_{-{1\over 2}} &=&-\partial_r \nonumber \\
M_0 &=& -\partial_{\zeta} \nonumber \\
Y_{1\over 2} &=& -t\partial_r-r\partial_{\zeta } \nonumber \\
X_0 &=& -t\partial_t-{1\over 2}r\partial_r-yg\partial_g-{x\over 2} \nonumber \\
X_1 &=& -t^2\partial_t-tr\partial_r-{1\over 2}r^2\partial_{\zeta}
-p_{01}tg\partial_g-xt \nonumber \\
N &=& -t\partial_t+\zeta\partial_{\zeta}-k_0g\partial_g \label{eq:rnageun} 
\EEA

{\bf Remark 2:} The case $x={1\over 2}$ is obtained from the above 
realization by setting $p_{01}=0$.

The MMG-representations of $\wit{\mathfrak{age}_1}$ which leave
invariant the linear Schr\"odinger equation co\"{\i}ncide with
eq.~(\ref{eq:nagdeux}). For $x=1/2$ there is no
restriction on $\Psi$ while the condition (\ref{eq:nerest}) must be satisfied
if $x\ne 1/2$. 

We shall need later that for the NMG-representation the linear
Schr\"odinger equation reads simply 
\BEQ
\left(2\partial_{\zeta }\partial_t
-\partial^2_r\right)\Psi(t,r,\zeta ,g)=0 \label{eq:ENMG}
\EEQ
while for the MMG-representation there arises an additional term
\BEQ
\left(2\partial_{\zeta }\partial_t-{4y\over \zeta} g\partial_g\partial_t-\partial^2_r\right)\Psi(t,r,\zeta ,g)=0 \label{eq:EMMG}
\EEQ
This follows from the explicit representations and the form (\ref{eq:cassch})
of the Schr\"odinger operator $\hat{S}$. 

\begin{table}
\caption[Represent 1]{Representations of the almost-parabolic and 
parabolic subalgebras  
$\mathfrak{g}\subset (\mathfrak{conf}_3)_{\mathbb{C}}$ and their 
(conditionally) invariant linear Schr\"odinger equation $\hat{S}\Psi=0$. 
The form of $\hat{S}$ and the auxiliary conditions only depend on the value
of $x$ but are independent of whether a parabolic or an almost-parabolic
subalgebra is considered. 
\label{tab1}}
\begin{tabular}{|||l|l||l|l|l|l|||} \hline\hline\hline
case & $\mathfrak{g}$ & representation & $x$ & auxiliary conditions & $\hat{S}$ \\ \hline\hline\hline
0 & $\mathfrak{age}_1$ & NMG & $=1/2$ & & $2\partial_{\zeta}\partial_t-\partial_r^2$                              \\
   & & $L=0$, $Q=0$, & & &   \\
  & & $P=p_{01}m_0 g^{(y+1)/y}$ & & &   \\ \cline{3-6}
  & & NMG & $\ne 1/2$ & see eq.~(\ref{age0condition}) &
            $2\partial_{\zeta}\partial_t-\partial_r^2$  \\
  & & $L=0$, $Q=0$,   & & & \\
  & & $P=p_{01}t^{y+1}m(t^y/g)$ & & 
  $\partial_{\zeta}\Psi=0$ & $\partial_r^2$ \\ \hline 
1 & $\mathfrak{age}_1$ & NMG & $=1/2$ & & $2\partial_{\zeta}\partial_t-\partial_r^2$ \\
\cline{4-6} 
  & & $L=0$, $Q=0$, & $\ne 1/2$ & $(2p_{01}g\partial_g+(2x-1))\Psi=0$ &
      $2\partial_{\zeta}\partial_t-\partial_r^2$ \\
  & & $P=p_{01}tg$ & & $\partial_{\zeta}\Psi=0$ & $\partial_r^2$ \\
\cline{1-1}\cline{3-3} 
2 & $\wit{\mathfrak{age}}_1$ & $K=k_0 g$ & & & \\ \hline \hline
3 & $\mathfrak{age}_1$ & MMG & $=1/2$ & & $2\partial_{\zeta}\partial_t-4yg\zeta^{-1}\partial_g\partial_{t}-\partial_r^2$ \\ \cline{4-6}
  & & $L=-2y\,g/\zeta$  & & &  \\
  & & $Q=-2y\,gr/\zeta$ & $\ne 1/2$ &
      $(\partial_{\zeta}-2y(g/\zeta)\partial_g)\Psi=0$ 
  & $(2\partial_{\zeta}\partial_t-\partial_r^2)\Psi=0$ \\
  & & $P=-ygr^2/\zeta$ & & & \\ \cline{1-1}\cline{3-3} 
4 & $\wit{\mathfrak{age}}_1$ & $K=k_0 g$ & & & \\ \hline\hline 
5 & $\mathfrak{alt}_1$ & $L=sg/\zeta$, $Q=srg/\zeta$ & $=1/2$ & & $2\partial_\zeta\partial_t+2sg\zeta^{-1}\partial_g\partial_t-\partial_r^2$ \\
  & & $P=sr^2g/2\zeta$ & & & \\ \cline{4-6}
  & & $F=2srg$ & $\ne 1/2$ &
      $(\partial_{\zeta}+sg\zeta^{-1}\partial_g)\Psi=\partial_r\Psi=0$ &
      $\partial_{\zeta}\partial_t$ \\ \cline{1-1}\cline{3-3}
6 & $\wit{\mathfrak{alt}}_1$ & $K=k_{0}' g$ & & & \\ \hline\hline
7 & $\mathfrak{sch}_1$ & $L=Q=P=0$ & $=1/2$ & $\partial_g\Psi=0$ & $2\partial_{\zeta}\partial_t-\partial_r^2$ \\ 
  & & & & $\partial_{\zeta}\Psi =0$ & $\partial_r^2$ \\ \cline{3-6}
  & & $L=Q=0$, $P=2ytg$ & $\ne 1/2$ & $\partial_{\zeta}\Psi=0$ 
  & $\partial_r^2$ \\ 
  & & & & $(4yg\partial_g+(2x-1))\Psi=0$ &
          $2\partial_{\zeta}\partial_t-\partial_r^2$ \\ 
\cline{1-1}\cline{3-3}
8 & $\wit{\mathfrak{sch}}_1$ & $K=k_0g$ & & & \\ \hline\hline\hline
\end{tabular}
\end{table}

We summarize the results of the classification of this and the following 
subsections in table~\ref{tab1}. For $\mathfrak{age}_1$ 
and $\wit{\mathfrak{age}}_1$ we distinguish five cases. Case 1 is obtained from case 0 by setting $m(v)=v^{-1}$ and only then there is an extension from
$\mathfrak{age}_1$ to $\wit{\mathfrak{age}}_1$ which is case 2. 
The cases 0, 1 and 3 refer to $\mathfrak{age}_1$ and distinguish between 
the NMG and MMG  representations as characterized by the explicit 
functions $L,Q,P$. For each of them, we give for both $x=1/2$ and 
$x\ne 1/2$ the explict form of the linear Schr\"odinger operator $\hat{S}$. 
For $x\ne 1/2$ there may be one or several auxiliary conditions which have to be met as well and in each case will lead to a modified form of the linear 
Schr\"odinger operator $\hat{S}$. In these cases, we have hence not found a Lie symmetry but rather a non-classical one, usually referred to as
$Q$-conditonal symmetry \cite{Blum69,Fush93}. 
Finally, the cases 2 and 4 apply to 
$\wit{\mathfrak{age}}_1$. Here the only new information is the explicit
form of $K$ whereas the form of $\hat{S}$ and the auxiliary conditions
remain unchanged.  
  
As an example, consider case 2. It refers to the NMG-representation of
$\wit{\mathfrak{age}}_1$ and the four functions $P,Q,L,K$ can be read off. 
Furthermore, one sees that for $x=1/2$, there is no further condition on
$\Psi$  and the linear Schr\"odinger operator is
$\hat{S}=2\partial_{\zeta}\partial_t-\partial_r^2$. On the other hand, for
$x\ne 1/2$, there are two {\em distinct} auxiliary conditions. For each of 
them, one reads off the corresponding invariant linear Schr\"odinger 
operator $\hat{S}$. The entry for case 1 can be read in the same way but
the function $K$ is of course not specified.  The other cases are understood
similarly.

\subsection{Subalgebras ${\mathfrak{alt}}_1$ and $\wit{\mathfrak{alt}}_1$}

For these algebras we fix the generators of dilatations $D$ and space 
translations $Y_{-{1/2}}$:
\BEA
Y_{-{1/2}} &=&-\partial_r \nonumber \\
D &=&-t\partial_t-r\partial_r-\zeta \partial_{\zeta}-sg\partial_g-x 
\label{eq:incon}
\EEA
where the exponent $s$ describes the scaling behaviour of the coupling $g$. 
The remaining generators are taken in the following general form
\BEA
M_0 &=&-\partial_{\zeta}-L(t,r,\zeta ,g)\partial_g \nonumber \\
Y_{1/2} &=& -t\partial_r-r\partial_{\zeta }
                 -Q(t,r,\zeta ,g)\partial_g \nonumber \\
X_1 &=& -t^2\partial_t-tr\partial_r-{1\over 2}r^2\partial_{\zeta}
                 -P(t,r,\zeta ,g)\partial_g-xt \nonumber \\
N &=& -t\partial_t+\zeta \partial_{\zeta }
      -K(t,r,\zeta ,g)\partial_g \nonumber \\
V_+ &=& -2tr\partial_t-2\zeta r\partial_{\zeta }-(r^2+2\zeta t)\partial_r
        -F(t,r,\zeta ,g)\partial_g-2xr 
\label{eq:ggen}
\EEA
Since this is a subalgebra of $(\mathfrak{conf}_3)_{\mathbb{C}}$,  
we obtain the following results for ${\mathfrak{alt}}_1$.
The comutators
\BEQ
{}[D,X_1]\:=\:-X_1\;\;,\;\;[D,Y_{1/2}]\:=\:0\;\;,\;\;
{}[D,M_0]\:=\:M_0\;\;,\;\;[D,V_+]\:=\:-V_+
\EEQ 
give
\BEQ
P=t^{s+1}p(u,v,w)\;\;,\;\; Q=t^sq(u,v,w)\;\;,\;\; 
L=t^{s-1}l(u,v,w) \;\;,\;\;F=t^{s+1}f(u,v,w)
\EEQ
where 
\BEQ
u={r\over t}\;\;,\;\;v=t^{-s}g\;\;,\;\;w={\zeta\over t} 
\EEQ
Next we use
\BEQ
{}[X_1,Y_{-{1\over 2}}]=Y_{1\over 2}\;\;,\;\;
{}[Y_{1\over 2},Y_{-{1\over 2}}]=M_0\;\;,\;\;
{}[Y_{-{1\over 2}},M_0]=0 \;\;,\;\;
{}[Y_{1\over 2},M_0]=0\;\;,\;\;[V_+,Y_{-{1\over 2}}]=2D
\EEQ
and find
\BEA
&&q(u,v,w)=u\partial_up(u,v,w)\;\;,\;\;l(u,v,w)=u\partial_u q(u,v,w)\;\;,\;\; 
\partial_ul(u,v,w,)=0 \nonumber \\
&&q(u,v,w)=ul(v,w)+n(v,w)\;\;,\;\;p(u,v,w)={u^2\over 2}l(v,w)+un(v,w)+m(v,w) 
\;\;,\;\; \nonumber \\
&&f(u,v,w)=4suv+z(v,w).
\EEA
hence $l=l(v,w)$. The remaining commutators
\BEA
{}[X_1,Y_{1\over 2}]=0\;\;,\;\;[V_+,M_0]=2Y_{1\over 2}\;\;,\;\;
{}[V_+,Y_{1\over 2}]=2X_1\;\;,\;\;[V_+,X_1]=0 
\EEA
give the following system for the unknown functions
$l(v,w),n(v,w),m(v,w)$ and $z(v,w)$
\BEA
\partial_wn+l\partial_vn-n\partial_vl &=& 0 \nonumber \\
(s-1)l-sv\partial_vl-w\partial_wl-\partial_wm+m\partial_vl-l\partial_vm&=&0 
\nonumber \\
(s-1)n-sv\partial_vn-w\partial_wn+m\partial_vn-n\partial_vm&=&0 \nonumber \\
2n+z\partial_vl-l\partial_vz&=&0 \nonumber \\
2n+z\partial_vl-l\partial_vz-\partial_wz&=&0 \nonumber \\
2m+2wl-2sv+z\partial_vn-n\partial_vz&=&0 \nonumber \\
2wn-(s+1)z+sv\partial_vz+w\partial_wz+z\partial_vn-n\partial_vz&=&0 
\EEA
The solution is readily found
\BEQ\label{eq:solalt}
z=0\;\;,\;\;n=0\;\;,\;\;l(v,w)=vl_1(w)\;\;,\;\;m(v,w)=sv-wvl_1(w)
\EEQ
and we arrive at the 
most general representation of the algebra $\mathfrak{alt}_1$ 
\BEA
D &=&- t\partial_t-r\partial_r-\zeta\partial_{\zeta}-sg\partial_g-x \nonumber \\
Y_{-{1\over 2}} &=& -\partial_r \nonumber \\
M_0 &=& -\partial_{\zeta }-t^{-1}l_1(w)g\partial_g \nonumber \\
Y_{1\over 2} &=& -t\partial_r-r\partial_{\zeta }
                 -t^{-1}rl_1(w)g\partial_g \nonumber \\
X_1 &=& -t^2\partial_t-tr\partial_r-{1\over 2}r^2\partial_{\zeta }
        -\left({t^{-1}r^2\over 2}l_1(w)+t(s-wl_1(w))\right)g\partial_g-xt 
\nonumber \\
V_+ &=& -2tr\partial_t-2\zeta r\partial_{\zeta }
        -(r^2+2\zeta t)\partial_r-2srg\partial_g-2xr 
\label{eq:galt}
\EEA
which depends on the parameters $x,s$ and arbitrary function $l_1(w)$.

For the algebra $\wit{\mathfrak{alt}_1}$ one must satisfy
also the commutators
\BEA
&&[D,N]=0\;\;,\;\;[Y_{-{1\over 2}},N]=0;\;,\;\;[M_0,N]=M_0\nonumber \\
&&[Y_{1\over 2},N]=Y_{1\over 2}\;\;,\;\;[X_1,N]=X_1\;\;,\;\;[V_+,N]=0
\EEA 
which gives $K=t^{s}k(u,v,w)$ and $\partial_uk(u,v,w)=0$ which means that
$k(u,v,w)=k(v,w)$ and furthermore we have the set of equations
\BEA
(2-s)l+sv\partial_vl+2w\partial_wl+\partial_wk+l\partial_vk-k\partial_vl&=&0 
\nonumber \\
(1-s)n+sv\partial_vn+2w\partial_wn+n\partial_vk-k\partial_vn&=&0 
\nonumber \\
sk-sv\partial_vk-w\partial_wk-sm+sv\partial_vm+2w\partial_wm+m\partial_vk
-k\partial_vm&=& 0 \nonumber \\
(s+1)z-sv\partial_vz-2w\partial_wz+z\partial_vk-k\partial_vz&=&0 
\label{eq:dkets}
\EEA 
Using eq.~(\ref{eq:solalt}) the system (\ref{eq:dkets}) reduces to the
single equation 
\BEQ
2vl_1+2vw\partial_wl_1+\partial_wk-kl_1+vl_1\partial_vk=0 \label{eq:soka}
\EEQ
and if we recall that $D=2X_0-N$ we obtain the final result
\BEQ\label{eq:fres}
k(v,w)=k'_0v\;\;,\;\;k'_0+s=2y\;\;;\;\;l_1(w)=l_0w^{-1}\;\;,\;\;K=k'_0g
\EEQ
where $k'_0$ and $l_0$ are constants. 
We therefore have the following realization of $\wit{\mathfrak{alt}}_1$
\BEA
D &=& -t\partial_t-r\partial_r-\zeta\partial_{\zeta}-sg\partial_g-x 
\nonumber \\
Y_{-{1\over 2}} &=& -\partial_r \nonumber \\
M_0 &=& -\partial_{\zeta }-{l_0\over \zeta}g\partial_g\nonumber \\
Y_{1\over 2} &=& -t\partial_r-r\partial_{\zeta }
                 -{l_0\over \zeta }rg\partial_g \nonumber \\
X_1 &=& -t^2\partial_t-tr\partial_r-{1\over 2}r^2\partial_{\zeta }
-\left({l_0\over 2\zeta}r^2+(s-l_0)t\right)g\partial_g-xt \nonumber \\
V_+ &=& -2tr\partial_t-2\zeta r\partial_{\zeta }
        -(r^2+2\zeta t)\partial_r-2srg\partial_g-2xr \nonumber \\
N &=&-t\partial_t+\zeta\partial_{\zeta}-k'_0g\partial_q \label{eq:nalt}
\EEA

Next we impose the invariance of the linear Schr\"odinger equation.
We have
\BEA
{}[\hat S,X_0] &=& -\hat S \nonumber \\
{}[\hat S,M_0] &=& [\hat S,Y_{-{1\over 2}}]\:=\:[\hat S,Y_{1\over 2}]\:=\:0 
\nonumber \\
{}[\hat S,X_1] &=& -2t\hat S -(1-2x)M_0 \nonumber \\
{}[\hat S, V_+] &=& -4r\hat S+2(1-2x)Y_{-{1\over 2}} \label{eq:inal}
\EEA
The conditions (\ref{eq:schinv}) are satisfied if $l_0=s$.

{\bf Remark 3:} If the generator $X_1$ is left out from 
$\mathfrak{alt}_1$ and $\wit{\mathfrak{alt}_1}$, the equation $\hat{S}\Psi=0$
is invariant even for arbitrary $l_0$.

In the case with fixed scaling dimension $x={1\over 2}$ the following
realization of algebra $\mathfrak{alt}_1$ is a dynamical symmetry of
the linear Schr\"odinger equation
\BEA
D &=& -t\partial_t-r\partial_r-\zeta\partial_{\zeta}-sg\partial_g-{1\over 2} \nonumber \\
Y_{-{1\over 2}} &=& -\partial_r \nonumber \\
M_0 &=& -\partial_{\zeta }-{s\over \zeta}g\partial_g\nonumber \\
Y_{1\over 2} &=& -t\partial_r-r\partial_{\zeta }-{s\over \zeta }rg\partial_g 
\nonumber \\
X_1 &=& -t^2\partial_t-tr\partial_r-{1\over 2}r^2\partial_{\zeta }
        -{s\over 2\zeta}r^2g\partial_g-{1\over 2}t \nonumber \\
V_+ &=& -2tr\partial_t-2\zeta r\partial_{\zeta }-(r^2+2\zeta t)\partial_r
        -2syrg\partial_g-2xr
\label{eq:lshal}
\EEA
without any restrictions on the wave function.

For arbitrary $x$ the conditions (\ref{eq:inal}) lead to 
very strong restrictions on the wave function
\BEA
M_0\Psi(\zeta,t,r,g)&=& 
(\partial_{\zeta }+{s\over \zeta}g\partial_g)\Psi(\zeta,t,r,g)=0 \nonumber \\
Y_{-1/2}\Psi(\zeta,t,r,g)&=&-\partial_r\Psi(\zeta,t,r,g)=0 
\label{stcon}
\EEA
which means that $\Psi(\zeta,t,r,g)=\psi(t,g^{1/s}\zeta)$. 

Finally, we add the generator $N$ in order to obtain the representation of 
$\wit{\mathfrak{alt}_1}$.  It is of the form 
$N=-t\partial_t+\zeta\partial_{\zeta}-k'_0g\partial_g$ and the
condition $[\hat S,N]=0$ is satisfied automatically.
Note that for this algebra the linear Schr\"odinger equation has the form:
\BEQ
\left(2\partial_{\zeta }\partial_t+{2s\over \zeta }g\partial_g\partial_t
-\partial^2_r\right)\Psi(\zeta,t,r,g)=0 \label{eq:AMMG}
\EEQ
These results are also included in table~\ref{tab1}.

\subsection{Subalgebras ${\mathfrak{sch}}_1$ and $\wit{\mathfrak{sch}}_1$}

In these algebras we must add to $\mathfrak{age}_1$ and 
$\wit{\mathfrak{age}}_1$ the generator of time
translation $X_{-1}=-\partial_t$ and satisfy
the commutator $[X_1,X_{-1}]=2X_0$. Inserting this condition into
the generators eq.~(\ref{eq:rageun}) we find 
$p_{01}=2y$ and $m(v)=v^{-1}=g t^{-y}$. 

The representation of ${\mathfrak{sch}}_1$ is 
\BEA
X_{-1} &=& -\partial_t \;\; , \;\; Y_{-{1\over 2}\:}=\:-\partial_r
\nonumber \\
M_0 &=& -\partial_{\zeta} \nonumber \\
Y_{1\over 2} &=& -t\partial_r-r\partial_{\zeta } \nonumber \\
X_0 &=& -t\partial_t-{1\over 2}r\partial_r-yg\partial_g-{x\over 2} 
\nonumber \\
X_1 &=& -t^2\partial_t-tr\partial_r-{1\over 2}r^2\partial_{\zeta}
          -2ytg\partial_g-xt 
\label{eq:rash} 
\EEA
and for $\wit{\mathfrak{sch}}_1$ we have $K=k_0g$, hence (\ref{eq:rash}) 
holds true, together with 
\BEQ
N = -t\partial_t+\zeta\partial_{\zeta}-k_0g\partial_g 
\label{eq:rnash} 
\EEQ
For the invariance of the linear Schr\"odinger equation we merely have
to consider a single commutator 
\BEQ
{}[\hat S,X_1]=-2t\hat S -(1-2x)M_0 +M_0{\hat Q}_{sch}\;\;,\;\; 
{\hat Q}_{sch}=4yg\partial_g . \label{eq:insh}
\EEQ
The linear Schr\"odinger equation is invariant if
\BEQ
M_0\Psi =0\;\;,\;\;\Psi(\zeta,t,r,g)=\psi(t,r,g)
\EEQ
which means that either the wave function does not depend on $\zeta $
or else
\BEQ
{\hat Q}_{sch}\Psi = (1-2x)\Psi \;\;,\;\;
\Psi(\zeta,t,r,g)=g^{(1-2x)/4y}\psi(\zeta,t,r)
\EEQ
Even in the case $x={1\over 2}$ we must impose an auxiliary condition 
on the wave functions. 

Again, we include our results in table~\ref{tab1}.

\section{Invariant nonlinear equations}

Having classified in the previous section the representations of the 
parabolic and almost-parabolic subalgebras of 
$(\mathfrak{conf}_3)_{\mathbb{C}}$ which leave the linear Schr\"odinger
equation $\hat{S}\Psi=0$ invariant, we now construct systematically
all semilinear invariant equations $\hat{S}\Psi=F(\zeta,t,r,g;\Psi,\Psi^*)$,
along the lines recalled in section~2.1. 

\subsection{Subalgebras $\mathfrak{age}_1$ and $\wit{\mathfrak{age}}_1$}

First we take the almost-parabolic subalgebra (\ref{eq:ageun}).
The potential $F$ must satisfy the system, see eq.(\ref{eq:repot})
\BEA
\partial_r F \:=\: 0\;\;,\;\;\partial_{\zeta}F &=&0 \nonumber \\
(2t\partial_t+2yg\partial_g
-x(\Psi \partial_{\Psi }+\Psi^*\partial_{\Psi^*})+(x+2))F &=& 0 
\nonumber \\
(t\partial_t+p_{01}t^ym(v)\partial_g
-x(\Psi \partial_{\Psi }+\Psi^*\partial_{\Psi^*})+(x+2))F &=& 0 
\EEA 
In the general case the solution is:
\BEQ
F^0_{\mathfrak{age}_1}=\Psi^{x+2\over x}f_0\left(\ln\Psi +\int{\!\D v\over v}{p_{01}vm(v)-2y\over p_{01}vm(v)-y},{\Psi\over \Psi^*}\right)
\EEQ
In the special case $m(v)=v^{-k}$ this leads to the following
form of $f_0$
\BEQ
f_0=f_0\left(\ln\left[t^{x/2}g^{-x/(2y)}(p_{01}-yt^{(k-1)y}g^{1-k})^{x\over 2y(k-1)}\Psi\right],{\Psi\over \Psi^*}\right)
\EEQ
Note that the case $x=1/2$ can be obtained by putting $k=(y+1)/y$. 

When the operator $\hat Q$ is taken in form (\ref{eq:ansatz}) the solution in 
the NMG-representation is
\BEQ
F^1_{\mathfrak{age}_1}=(t^{p_{01}-2y}g)^{-{x+2\over 2(p_{01}-y)}}
f^1_{\mathfrak{age}_1}\left((t^{p_{01}-2y}g)^
{x\over 2(p_{01}-y)}\Psi ,{\Psi\over {\Psi}^*}\right)\label{eq:ping}
\EEQ
So, the nonlinear Schr\"odinger equation is
\BEQ
(2\partial_{\zeta }\partial_t-\partial^2_r)\Psi(\zeta,t,r,g)=
F^1_{\mathfrak{age}_1} \label{eq:nplag}
\EEQ

For the extention to the parabolic algebra $\wit{\mathfrak{age}}_1$, 
we must add the equation
\BEQ
(t\partial_t+k_0g\partial_g)F=0
\EEQ
which leads to the following equation
for $f^1_{\mathfrak{age}_1}$
\BEQ
{p_{01}-2y+k_0\over 2(p_{01}-y)}
\left(x+2-xu\frac{\partial}{\partial u}\right)f^1_{\mathfrak{age}_1}(u,v)=0.
\EEQ
where $u=(t^{p_{01}-y}g)^{x\over 2(p_{01}-y)}\Psi$ and  $v=\Psi/\Psi^*$.

There are two cases:
\begin{enumerate}
\item the generic solution $p_{01}\ne 2y-k_0$
\BEQ
F^1_{\wit{\mathfrak{age}}_1}=\Psi^{x+2\over x}
f_{\mathfrak{sch}_1}\left({\Psi\over {\Psi}^*}\right)
\label{eq:shnl}
\EEQ
which is the same as for the nonmodified representation of the 
Schr\"odinger algebra.
\item a non-generic solution $p_{01}=2y-k_0$. Here the form of the 
potential is the same as in (\ref{eq:ping}) but one has an additional
constraint on the parameters of the algebra. 
\end{enumerate}
Note that in the canonical case $x={1\over 2}$ one must also put $p_{01}=0$
in the above results.

We consider now the MMG-realisation (\ref{eq:rnagdeux}) of the algebra $\mathfrak{age}_1$. The potential is found from 
\BEA
\partial_r F=0\;\;,\;\;
\left(\partial_{\zeta}-2y\frac{g}{\zeta}\partial_g\right)F &=&0 \nonumber \\
(2t\partial_t+2yg\partial_g-x(\Psi \partial_{\Psi }+\Psi^*\partial_{\Psi^*})+(x+2))F &=&0 \nonumber \\
(t\partial_t-x(\Psi \partial_{\Psi }+\Psi^*\partial_{\Psi^*})+(x+2))F&=&0 
\EEA 
which has the solution
\BEQ\label{eq:gpot}
F^2_{\mathfrak{age}_1}=b^{x+2}
f^2_{\mathfrak{age}_1}\left(b^{-x}\Psi ,{\Psi\over \Psi^*}\right)
\;\;,\;\; b=t^{-1}\zeta g^{1/(2y)}
\EEQ
The nonlinear equation invariant under the same algebra is
\BEQ
\left(2\partial_{\zeta }\partial_t-{4y\over \zeta }g\partial_g-\partial^2_r\right)\Psi(\zeta,t,r,g)=F^2_{\mathfrak{age}_1} \label{eq:nalge}
\EEQ

When we extend this to the parabolic subalgebra $\wit{\mathfrak{age}}_1$ by
including the generator $N$ (see (\ref{eq:nagdeux})),
we have
\BEQ
\left({k_0\over 2y}-2\right)
\left(x+2-xu\frac{\partial}{\partial u}\right)f^2_{\mathfrak{age}_1}(u,v)=0.
\EEQ
where $u=b^{x}\Psi $ and $v=\Psi/\Psi^*$. Our results give
\begin{enumerate}
\item the generic solution $k_0\ne 4y$.
We recover the same result as for 
nonmodified Schr\"odinger algebra (\ref{eq:shnl}).
\item a non-generic solution $k_0=4y$. The result is the same
as in (\ref{eq:gpot}), but the central generator $N=-t\partial_t+\zeta\partial_{\zeta}-4yg\partial_g$ is
specified.
\end{enumerate}

\noindent
These results are valid for arbitrary scaling dimension, but
for the case $x={1\over 2}$ there is no restriction on the wave functions.

\begin{table}
\caption{Solutions for the invariant semilinear potential of the 
equation $\hat{S}\Psi=F$. The cases are the ones from table~\ref{tab1} and
$f$ is a generic symbol for an arbitrary function. \label{tab2}}
\begin{tabular}{||l|l||l|l|l||} \hline \hline
case & subalgebra  & potential $F$ & condition & \\ \hline \hline
0 & $\mathfrak{age}_1$ & $\Psi^{x+2\over x}f\left(\ln\Psi +\int{\!\D v\over v}{p_{01}vm(v)-2y\over p_{01}vm(v)-y},\Psi/\Psi^*\right)$ 
& $m(v)$ arbitrary & $v=t^y/g$ \\ \hline
1 & $\mathfrak{age}_1$ & $a^{x+2}f(a^x\Psi, \Psi/\Psi^*)$ & &  $a=[t^{p_{01}-2y}g]^{1/(2(p_{01}-y))}$ \\
2 & $\wit{\mathfrak{age}}_1$ & $\Psi^{(x+2)/x}f(\Psi/\Psi^*)$ & $p_{01}\ne2y-k_0$ & \\ 
  & $\wit{\mathfrak{age}}_1$  & $a^{x+2}f(a^x\Psi, \Psi/\Psi^*)$ &
    $p_{01}=2y-k_0$ & $a=[gt^{-k_0}]^{1/2(y-k_0)}$ \\ \hline
3 & $\mathfrak{age}_1$ & $b^{(x+2)}f(b^{-x}\Psi,\Psi/\Psi^*)$ & & 
$b=t^{-1}\zeta g^{1/2y}$ \\
4 & $\wit{\mathfrak{age}}_1$ & $\Psi^{(x+2)/x}f(\Psi/\Psi^*)$ & $k_0\ne 4y$ 
& \\ 
  & $\wit{\mathfrak{age}}_1$ & $b^{(x+2)}f(b^{-x}\Psi,\Psi/\Psi^*)$ &
    $k_0=4y$ & $b=t^{-1}\zeta g^{1/2y}$ \\ \hline
5 & $\mathfrak{alt}_1$ & $t^{-x-2}f(\zeta^{-s}g,t^x\Psi,\Psi/\Psi^*)$ & &\\
6 & $\wit{\mathfrak{alt}}_1$ & $c^{-x-2}f(c^x\Psi,\Psi/\Psi^*)$ & &  $c=(\zeta^sg^{-1})^{1/(s+k'_0)}t$ \\ \hline
7 & $\mathfrak{sch}_1$ & $g^{-(x+2)/2y}f(g^{x/2y}\Psi,\Psi/\Psi^*)$ & & \\
8 & $\wit{\mathfrak{sch}}_1$ & $\Psi^{(x+2)/x}f(\Psi/\Psi^*)$ & $k_0\ne 0$ & \\ 
  & $\wit{\mathfrak{sch}}_1$ & $g^{-(x+2)/2y}f(g^{x/2y}\Psi,\Psi/\Psi^*)$ &
    $k_0=0$ & \\ \hline\hline
\end{tabular}
\end{table}

The results for the non-linear potential $F$ are collected in the 
table~\ref{tab2}. We give the 
generic solutions for the cases as defined in table~\ref{tab1}. 
For each of the cases, we had the linear equation $\hat{S}\Psi=0$
together with eventual auxiliary condition(s) for a given value of $x$. Now
the associated non-linear equation is simply $\hat{S}\Psi=F$, where $F$
is read from the table. In table~\ref{tab2} we also list the non-generic solutions which are distinguished by the conditions mentioned and for
some of which there are modified scaling variables to be used.

\subsection{Subalgebras $\mathfrak{alt}_1$ and $\wit{\mathfrak{alt}}_1$}

For the realization (\ref{eq:lshal}) of $\mathfrak{alt}_1$ 
(note that for this realization of the algebra one has 
$k'_0+s=2y$ by construction), the potential is found from 
\BEA
\partial_r F=0\;\;,\;\;(\zeta \partial_{\zeta}+sg\partial_g)F&=&0 \nonumber \\
(t\partial_t+\zeta \partial_{\zeta }+sg\partial_g-x(\Psi \partial_{\Psi }+\Psi^*\partial_{\Psi^*})+(x+2))F&=&0 \nonumber \\
(t\partial_t-x(\Psi \partial_{\Psi }+\Psi^*\partial_{\Psi^*})+(x+2))F&=&0 
\EEA 
to have the form
\BEQ\label {eq:tpot}
F_{\mathfrak{alt}_1}=t^{-x-2}
f_{\mathfrak{alt}_1}\left(t^x\Psi , \zeta^{-s}g, {\Psi\over \Psi^*}\right) 
\EEQ
When we want to have also invariance under $\wit{\mathfrak{alt}}_1$
the $N$ operator gives 
$(t\partial_t-\zeta \partial_{\zeta}+k'_0g\partial_g)F_{\mathfrak{alt}_1}=0$.
So in this case we have
\BEQ\label{eq:ntpot}
F_{\wit{\mathfrak{alt}}_1}=a^{-x-2}
f_{\wit{\mathfrak{alt}}_1}\left(a^x\Psi ,{\Psi\over \Psi^*}\right)
\;\;,\;\;
a=t\zeta^{s/(s+k'_0)} g^{-{1/(s+k'_0)}}
\EEQ
The nonlinear equation invariant under the same algebra is
\BEQ
\left(2\partial_{\zeta }\partial_t+4y{g\over \zeta}\partial_g\partial_t
-\partial^2_r\right)\Psi(\zeta,t,r,g)=
F_{\mathfrak{alt}_1}(F_{\wit{\mathfrak{alt}}_1})\label{eq:nlalt}
\EEQ
and the results are again included in table~\ref{tab2}. We point
out that the form of $F$ agrees with the special solution obtained
before for  $\wit{\mathfrak{age}}_1$-MMG representation.

\subsection{Subalgebras $\mathfrak{sch}_1$ and $\wit{\mathfrak{sch}_1}$}

Now the system is
\BEA
\partial_r F=0\;\;,\;\;\partial_{\zeta}F=0 \;\;,\;\;\partial_t F&=&0 
\nonumber \\
(2yg\partial_g-x(\Psi \partial_{\Psi }+\Psi^*\partial_{\Psi^*})+(x+2))F&=&0
\EEA 
which has the solution
\BEQ\label{eq:shpot}
F_{\mathfrak{sch}_1}=g^{-{(x+2)/(2y)}}
f_{\mathfrak{sch}_1}\left(g^{x/(2y)}\Psi , {\Psi\over \Psi^*}\right) 
\EEQ
The result for the algebra  $\wit{\mathfrak{sch}_1}$ we obtain by
adding the invariance under generator $N$, which leads to the following
equation 
\BEQ
{k_0\over 2y}\left(x+2-xu\frac{\partial}{\partial u}\right)
f_{\mathfrak{sch}_1}(u,v)=0\;\;,\;\;u=g^{x/(2y)}\Psi\;\;,\;\; v=\Psi/\Psi^*
\EEQ
with solutions:

i) in the case $k_0\ne 0$  like (\ref{eq:shnl}).

ii) in the case $k_0=0$ like (\ref{eq:shpot}), but with a non-modified
central generator $N$.

\noindent 
The nonlinear equation has the generic form
\BEQ \label{eq:nsche}
(2\partial_{\zeta }\partial_t-\partial^2_r)
\Psi(t,r,\zeta ,g) = F_{\mathfrak{sch}_1} 
\EEQ

\section{Consequences and conclusion}

Motivated by the problem to understand the form of the scaling functions of
the two-time observables in phase-ordering kinetics, we have been led to 
reconsider the question of finding semilinear Schr\"odinger equations which are 
invariant under a conveniently chosen representation 
of the Schr\"odinger group. 
The main difficulty is that if one considers the `mass' as a fixed constant,
Galilei- together with spatial-translation invariance implies that such an
equation should be invariant under simple phase shifts which severely restricts
the possible form of a non-linear potential and in general is only possible for
complex wave functions $\Phi$. As we have seen, a possible way out of this
difficulty is to consider the `mass' as an additional dynamcial variable
and then to go over to a `dual' formulation with a wave function
$\Psi=\Psi(\zeta,t,r)$, see eq.~(\ref{eq:Fourier}). In this way the 
Schr\"odinger algebra $\mathfrak{sch}_d$ is actually embedded into a
conformal algebra $(\mathfrak{conf}_{d+2})_{\mathbb{C}}$ which naturally leads
to the question of finding all semilinear Schr\"odinger equations 
$\hat{S}\Psi=F(\Psi,\Psi^*)$ conditionally invariant under some 
parabolic or almost-parabolic
subalgebra of $(\mathfrak{conf}_{d+2})_{\mathbb{C}}$, see figure 1bcd.  

We then considered two extensions by further giving up some other 
property which is habitually admitted:
\begin{enumerate}
\item non-hermitian representations were constructed in section~2 and the
corresponding non-linear equations are  found, 
see eqs.~(\ref{2:sch},\ref{2:age},\ref{2:alt1},\ref{2:alt2}). 
\item a dimensionful coupling $g$ was explicitly introduced into the 
potential. The classification of the differential operator representations
which also leave the linear equation $\hat{S}\Psi=0$ invariant 
is given in table~\ref{tab1} and the corresponding semi-linear 
Schr\"odinger equations $\hat{S}\Psi=F$ are listed in table~\ref{tab2}. 
\end{enumerate}
Besides this classification, we think the most remarkable result is that
quite generally, {\em real-valued} solutions of these invariant equations
are obtained. 

To illustrate the possible impact on the understanding of phase-ordering
kinetics, we reconsider 
eq.~(\ref{eq:ping}) with $p_{01}=2y$ or else eq.~(\ref{eq:shpot}) 
together with their auxiliary condition. We write the wave function as
$\Psi(\zeta,t,r,g) = g^{(1-2x)/(4y)}\psi(\zeta,t,r)$ where $\psi$ is 
a real-valued function which satisfies the equation
\BEQ
\left( 2\partial_{\zeta}\partial_t - \partial_r^2\right) \psi = g^{-5/(4y)}
f\left( g^{1/(4y)} \psi\right) = \psi^5 \bar{f}\left(g \psi^{4y}\right)
\EEQ
where $f$ is an arbitrary function and ${f}(\mathfrak{x})=\mathfrak{x}^5 \bar{f}(\mathfrak{x})$. Up to the Fourier/Laplace transform with 
respect to $\zeta$,
this is of the same form as the coarse-grained kinetic equation 
(\ref{modelA}) habitually used to describe coarsening. Since quite different
functions $f$ can be described in terms of the same symmetry, this might
provide an explanation for the well-established fact \cite{Bray94} that the
long-time behaviour of correlators and response functions in 
phase-ordering kinetics is quite independent of the
precise form of the potential $V(\Phi)$. We shall elaborate on this 
elsewhere.

On the other hand, one may use these symmetries to reduce the nonlinear Schr\"odinger equation
to a linear equation and hence obtain new explicit solutions, along the lines of \cite{Fein04}. We hope to return to this elsewhere.

\noindent 
{\bf Acknowledgements:} 
We thank R. Cherniha and R. Schott for useful conversations. 
S.S. was supported by the EU Research Training Network HPRN-CT-2002-00279.



\begin{thebibliography}{999}

\bibitem{Abri04a} S. Abriet and D. Karevski, Eur. Phys. J. {\bf B37}, 43 (2004).

\bibitem{Abri04b} S. Abriet and D. Karevski, Eur. Phys. J. {\bf B41}, 79 (2004).

\bibitem{Aran02} I.S. Aranson and L. Kramer, Rev. Mod. Phys. {\bf 74}, 
100 (2002). 

\bibitem{Baru80} A.O. Barut and R. Raczka, {\it Theory of group representations
and applications}, Polish Science Publications (Varsovie 1980). 

\bibitem{Baum05} F. Baumann, M. Henkel, M. Pleimling and J. Richert, submitted
to J. Phys. {\bf A} (2005); {\tt cond-mat/0504243}. 

\bibitem{Bazh04} V.V. Bazhanov, S.L. Lukyanov and A.B. Zamolodchikov, 
Adv. Theor. Math. Phys. {\bf 7}, 711 (2004). 

\bibitem{Blum69} G.W. Bluman and J.D. Cole, J. Math. Mech. {\bf 18}, 1025 
(1969).

\bibitem{Bouc00} J.P. Bouchaud, in M.E.Cates and M.R.Evans (Eds), {\it
Soft and Fragile Matter}, IOP Press, Bristol (2000).

\bibitem{Bour00} J. Bourgain, {\it Global solutions of non-linear Schr\"odinger
equations}, Am. Math. Society (1999). 

\bibitem{Boye76} C.D. Boyer, R.T. Sharp and P. Winternitz,
J. Math. Phys. {\bf 17}, 1439 (1976).


\bibitem{Bray94} A.J. Bray, Adv. Phys. {\bf 43}, 357 (1994).

\bibitem{Bray00} A.J. Bray, in M.E.Cates and M.R.Evans (Eds), {\it
Soft and Fragile Matter}, IOP Press, Bristol (2000).

\bibitem{Burd73} G. Burdet, M. Perrin and P. Sorba: Comm. Math. Phys. {\bf 34},
85 (1973);\\
G. Burdet, J. Patera, M. Perrin et P. Winternitz, {\it Lie subalgebras and Schr\"odinger algebra} (in French), preprint CRM-689 (f\'ev 1997).


\bibitem{Caud04} V. Caudrelier, M. Minchev and E. Ragoucy, J. Phys. 
{\bf A37}, L367 (2004). 

\bibitem{Cher00} R. Cherniha and J.R. King, J. Phys. {\bf A33}, 
267 and 7839 (2000). 

\bibitem{Cher03} R. Cherniha and J.R. King, J. Phys. {\bf A36}, 405 (2003). 

\bibitem{Cher04} R. Cherniha and M. Henkel, J. Math. Anal. Appl. {\bf 298},
487 (2004). 

\bibitem{Corb02} F. Corberi, E. Lippiello and M. Zannetti, Phys. Rev. 
{\bf E65}, 046136 (2002). 

\bibitem{Cris03} A. Crisanti and F. Ritort, J.Phys. {\bf A36}, R181 (2003).

\bibitem{Cugl94} L.F. Cugliandolo, J. Kurchan and G. Parisi, J. Physique
{\bf I4}, 1641 (1994). 

\bibitem{Cugl02} L.F. Cugliandolo, {\it Dynamics of Glassy Systems};
{\tt cond-mat/0210312}.

\bibitem{deMo99} M. de Montigny, F.C. Khanna, A.E. Santana, E.S. Santos and
J.D.M. Vianna, Ann. of Phys. {\bf 277}, 144 (1999). 

\bibitem{Dira63} P.A.M. Dirac, J. Math. Phys. {\bf 4}, 901 (1963). 

\bibitem{Dobr97} V.K. Dobrev, H.D. Doebner and C. Mrugalla, Reports Math. Phys. {\bf 39}, 201 (1997).

\bibitem{Dobr99} V.K. Dobrev, H.D. Doebner and C. Mrugalla, Mod. Phys.
Lett. {\bf A14}, 1113 (1999).

\bibitem{Doeb95} H.-D. Doebner and H.-J. Mann, J. Math. Phys. {\bf 36},
3210 (1995). 
 
\bibitem{Dorn01} I. Dornic, H. Chat\'e, J. Chave and H. Hinrichsen,
Phys. Rev. Lett. {\bf 87}, 045701 (2001). 

\bibitem{Dorn02} I. Dornic, {\it Th\`ese de doctorat}, (Nice et Saclay 2002).

\bibitem{Dunn89} G.V. Dunne, R. Jackiw and C.A. Trugenberger, Ann. of
Phys. {\bf 194}, 197 (1989). 

\bibitem{Fein04} P. Feinsilver, Y. Kocik and R. Schott, Fortschritte Physik {\bf 52}, 343 (2004). 

\bibitem{Flor94} R. Floreanini and L. Vinet, Lett. Math. Phys. {\bf 32},
37 (1994). 

\bibitem{Fron65} C. Fronsdal, Rev. Mod. Phys. {\bf 37}, 221 (1965); 
Phys. Rev. {\bf D10}, 589 (1974); {\bf D12}, 3819 (1975); {\bf D26}, 1988 (1982); M. Flato and C. Fronsdal, Phys. Lett. {\bf 97B}, 236 (1980). 

\bibitem{Fush93} W.I. Fushchich, W.M. Shtelen and N.I. Serov, {\it Symmetry
analysis and exact solutions of equations of nonlinear mathematical physics},
Kluwer (Dordrecht 1993). 

\bibitem{Fush95} W.I. Fushchich and R.M. Cherniha, J. Phys. {\bf A28}, 
5569 (1995). 

\bibitem{Godr00a} C. Godr\`eche and J.M. Luck, J. Phys. {\bf A33}, 1151 (2000).

\bibitem{Godr00b} C. Godr\`eche and J.M. Luck, J. Phys. {\bf A33}, 9141 (2000).

\bibitem{Godr02} C. Godr\'eche and J.M. Luck, J. Phys. Cond. Matt,
{\bf 14}, 1589 (2002).

\bibitem{Giul96} D. Giulini, Ann. of Phys. {\bf 249}, 222 (1996). 

\bibitem{Guen99} F. G\"ung\"or, J. Phys. {\bf A32}, 977 (1999). 

\bibitem{Hage72} C.R. Hagen,  Phys. Rev. {\bf D5}, 377 (1972). 

\bibitem{Hass00} M. Hassa\"{\i}ne and P.A. Horv\'athy, Ann. of Phys. {\bf 282},
218 (2000); Phys. Lett. {\bf A279}, 215 (2001). 

\bibitem{Henk94} M. Henkel, J. Stat. Phys. {\bf 75}, 1023 (1994). 

\bibitem{Henk01} M. Henkel, M. Pleimling, C. Godr\`eche and J.-M. Luck,
Phys. Rev. Lett. {\bf 87}, 265701 (2001).

\bibitem{Henk02} M. Henkel, Nucl. Phys. {\bf B641}, 405 (2002).

\bibitem{Henk03b} M. Henkel and J. Unterberger, 
Nucl. Phys. {\bf B660}, 407 (2003).

\bibitem{Henk03c} M. Henkel and M. Pleimling, Phys. Rev. {\bf E68}, 065101(R)
(2003). 

\bibitem{Henk03d} M. Henkel and G.M. Sch\"utz, J. Phys. {\bf A37}, 591 (2004).

\bibitem{Henk04} M. Henkel, Adv. Solid State Phys. {\bf 44}, 389 (2004). 

\bibitem{Henk04b} M. Henkel, A. Picone and M. Pleimling, Europhys. Lett. 
{\bf 68}, 191 (2004). 

\bibitem{Henk05} M. Henkel, {\tt cond-mat/0503739}. 

\bibitem{Hohe77} P.C. Hohenberg and B.I. Halperin, Rev. Mod. Phys. {\bf 49},
435 (1977).


\bibitem{Kac87} V.G. Kac and A.K. Raina, {\it Bombay lectures on heighest-weight
representations of infinite-dimensional Lie algebras}, World Scientific
(Singapour 1987). 

\bibitem{Kast68} H.A. Kastrup, Nucl. Phys. {\bf B7}, 545 (1968). 

\bibitem{Knap86} A.W. Knapp, {\it Representation theory of semisimple groups:
an overview based on examples}, Princeton University Press (Princeton 1986). 

\bibitem{Levi89} D. Levi and P. Winternitz, J. Phys. {\bf A22}, 2915 (1989). 

\bibitem{Levi05} D. Levi and P. Winternitz, {\tt nlin.SI/0502004}. 

\bibitem{Lipp00} E. Lippiello and M. Zannetti, Phys. Rev. {\bf E61}, 
3369 (2000).

\bibitem{Lore05} E. Lorenz and W. Janke, to be published (2005). 


\bibitem{Medi85} A. Medina et P. Revoy, Annales scient. \'ecole normale sup\'erieure, 4$^e$ s\'erie, {\bf 18}, 533 (1985). 

\bibitem{Nied72} U. Niederer, Helv. Phys. Acta {\bf 45}, 802 (1972). 

\bibitem{ORai01} L. O'Raifeartaigh and V.V. Sreedhar, Ann. of Phys. {\bf 293},
215 (2001). 

\bibitem{Perr77} M. Perroud, Helv. Phys. Acta {\bf 50}, 233 (1977). 

\bibitem{Pico02} A. Picone and M. Henkel, J. Phys. {\bf A35}, 5575 (2002).

\bibitem{Pico04} A. Picone and M. Henkel, Nucl. Phys. {\bf B688}, 217 (2004).

\bibitem{Popo04} R.O. Popovich, N.M. Ivanova and H. Eshraghi, J. Math. Phys.
{\bf 45}, 3049 (2004). 


\bibitem{Ride93} G. Rideau and P. Winternitz, J. Math. Phys. {\bf 34}, 558 (1993). 

\bibitem{Rute95} A.D. Rutenberg and A.J. Bray, Phys. Rev. {\bf E51}, 
5499 (1995).


\bibitem{Shat00} J. Shatah and M. Struwe, {\it Geometric wave equations}, Courant Lecture Notes in Mathematics vol 2, American Mathematical Society
(New York 2000). 


\bibitem{Sule99} C. Sulem and P.-L. Sulem, 
{\it The non-linear Schr\"odinger equation }, Appl. Math. Sci. {\bf 139}, Springer (1999). 


\bibitem{Zipp00} W. Zippold, R. K\"uhn and H. Horner, Eur. Phys. J. {\bf B13},
531 (2000). 

\end{thebibliography}
\end{document}